\documentclass[journal]{IEEEtran}

\usepackage{mathptmx}
\usepackage{multicol}
\usepackage{lipsum}
\usepackage{setspace}
\usepackage{xcolor}
\usepackage{wrapfig}
\usepackage{comment}
\usepackage[square,numbers]{natbib}
\usepackage{mathtools}
\usepackage{amssymb,amsmath,amsthm}
\newtheorem{theorem}{Theorem}
\usepackage {graphicx}
\usepackage{epsfig}
\usepackage{epstopdf}
\usepackage{caption}
\usepackage{longtable}
\usepackage{subcaption}

\usepackage{lineno,hyperref}

\usepackage{algorithm }
\usepackage{algpseudocode}

\usepackage{multirow}

\usepackage{fancyvrb}

\usepackage{mathptmx}
\usepackage{alltt}

\usepackage{pdflscape}
\usepackage{rotating}
\usepackage{lscape}

\usepackage{booktabs}

\usepackage{tabularx}
\usepackage{array}

\usepackage[english]{babel}
\usepackage{babel}

\algnewcommand\algorithmicforeach{\textbf{for each}}
\algdef{S}[FOR]{ForEach}[1]{\algorithmicforeach\ #1\ \algorithmicdo}

\begin{document}

\title{Cloud computing as a platform for monetizing data services: A two-sided game business model}

\author{Ahmed Saleh Bataineh,
        Jamal Bentahar,
        Rabeb Mizouni,
        Omar Abdel Wahab,
        Gaith Rjoub,
        May El Barachi

}

\markboth{Journal of \LaTeX\ Class Files,~Vol.~14, No.~8, August~2015}%
{Shell \MakeLowercase{\textit{et al.}}: Bare Demo of IEEEtran.cls for Computer Society Journals}

\maketitle

\IEEEtitleabstractindextext{%
\begin{abstract}
With the unprecedented reliance on cloud computing as the backbone for storing today's big data, we argue in this paper that the role of the cloud should be reshaped from being a passive virtual market to become an active platform for monetizing the big data through Artificial Intelligence (AI) services. The objective is to enable the cloud to be an active platform that can help big data service providers reach a wider set of customers and cloud users (i.e., data consumers) to be exposed to a larger and richer variety of data to run their data analytic tasks. To achieve this vision, we propose a novel game theoretical model, which consists of a mix of cooperative and competitive strategies. The players of the game are the big data service providers, cloud computing platform, and cloud users. The strategies of the players are modeled using the two-sided market theory that takes into consideration the network effects among involved parties, while integrating the externalities between the cloud resources and consumer demands into the design of the game. Simulations conducted using Amazon and google clustered data show that the proposed model improves the total surplus of all the involved parties in terms of cloud resources provision and monetary profits compared to the current merchant model.
\end{abstract}

}

\IEEEdisplaynontitleabstractindextext

\IEEEpeerreviewmaketitle

\section{Introduction}\label{sec:introduction}

\IEEEPARstart{C}{loud} computing is witnessing a striking increase in the number of enterprises and manufacturers that are relying on this paradigm to store and process their data. For example, the study reported in \citep{DMRAMAZON2018} revealed that one million customers deploy their own enterprises on Amazon, spending $30$ billion USD on persistent storage on Amazon EC$2$ instances and generating $600$ ZB of data per year \citep{Niyato2018}. This explosive amount of data generated and stored on cloud resources forms the backbone for Artificial Intelligence (AI) services and opens the door for a new cloud business paradigm, enabling the latter to be an active platform for monetizing data that benefit AI services. However, the cloud is not the actual owner for these big chunks of data, and has no right to trade and use these data without considering its actual owners.

\begin{figure*}[!htp]
\hspace{1.5cm}
    \includegraphics[width=\textwidth,height=12cm]{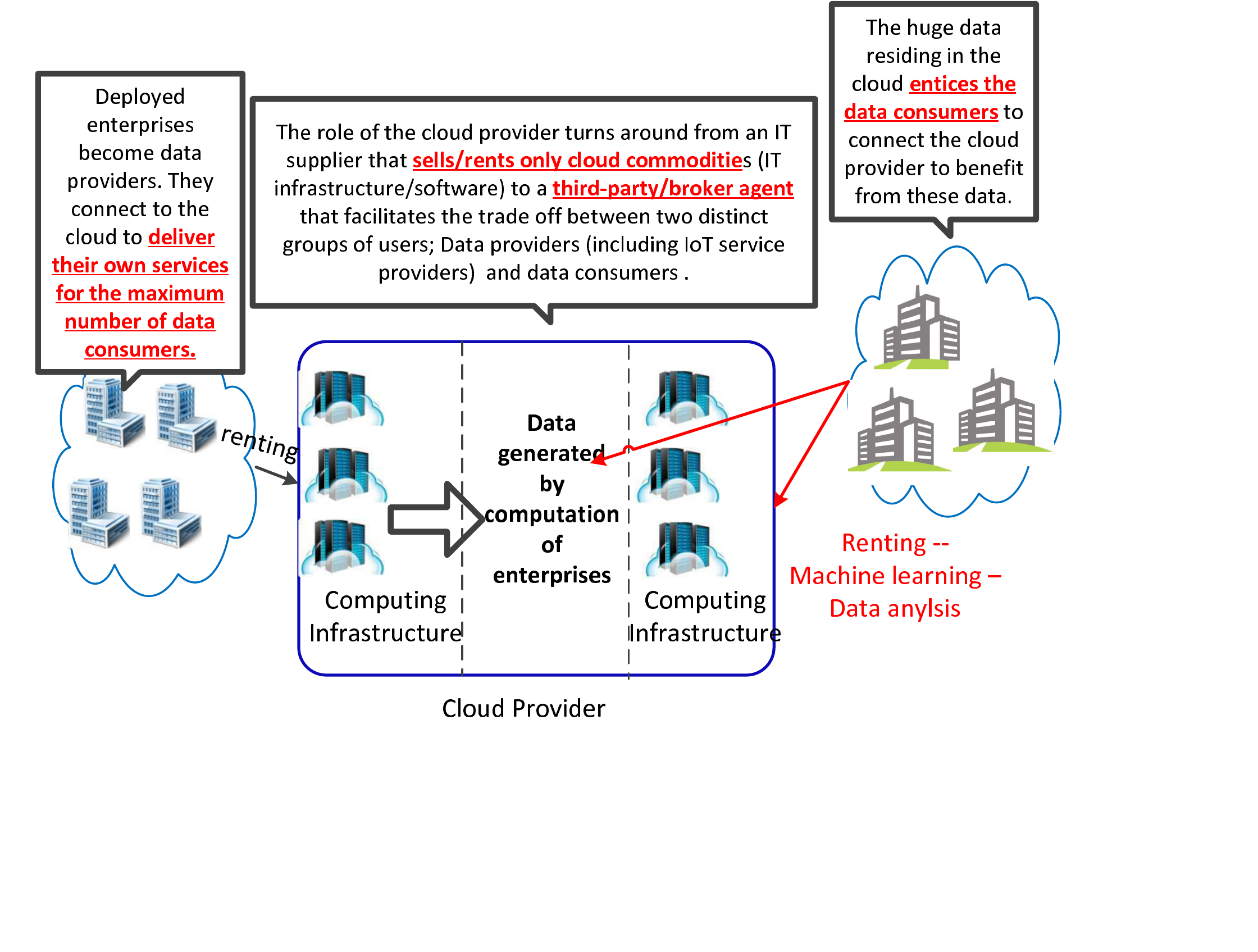}
    \vspace{- 3cm}
    \caption{Overview of the new cloud business model}
    \label{overview_idea}
\end{figure*}

Motivated by the vision of the cloud as platform for monetizing data services, we propose in this paper a novel cloud business model which allows data consumers (e.g., market research enterprises) to run their data analytics  on the huge and diverse data that are stored on the cloud. This not only gives data consumers the opportunity to extract valuable patterns from massive data coming from multiple data providers, but also releases them from having to search and discover appropriate providers for each particular type of data they need to analyze. Data providers, in addition to favoring the access to cloud-based infrastructure over purchasing their own computing and storage platforms, find in the enormous and varied number of data consumers that deal with the cloud an extra motivation to store their data on this platform to improve their exposure and increase their market shares. This indirectly makes, as shown in Figure \ref{overview_idea}, the cloud computing platform a mediator between data providers and data consumers and a principal player in the whole big data analytics process. This opens the door for new and innovative business models to take advantage of this scenario to increase the profits of all the involved parties, apart from the traditional business models which treat the cloud as being a passive virtual market for offering services via the Internet.
Specifically, the literature on business-oriented data trading can be classified into two main categories, i.e., pure merchant approaches and collaborative approaches. The proposals under the pure merchant approach such as \citep{Niyato2016, Niyato20166} and \citep{Niyato2018} adopt classic economic approaches, mainly the demand-supply model and one-sided game theory/auction-based pricing to model the interactions among data providers, data consumers and third-party platforms (i.e., information service providers). In this approach, the third-party platform aims to maximize its revenue through buying data from their owners, reprocessing them, extracting useful information and selling this information to consumers. This approach suffers from several limitations when applied in cloud computing scenarios. The first limitation is related to the diversity in the data consumers’ interests, which entails higher processing costs (in terms of information extraction for different customers' interests) for the third-party platform. Moreover, under the pure merchant model, data providers aim to maximize their revenue of using their data commodities while the third-party platform aims to minimize the cost of raw data bought from these providers. In parallel, from the data consumers' side, the third-party platform aims to maximize its revenue from selling the processed information while consumers aim to minimize the cost of information commodities, considering the maximum available quality and quantity of information. The resulting equilibrium from such aggressive competitions among the different involved parties leads to less and coarse distribution of the total surplus. In addition, data differ from other economic goods for its potential of being (re)-sold to many consumers at the same time. In the pure merchant model, since economic goods cannot be resold, the equilibrium of the market lies at the intersection of the demand and supply curves. This means that the quantities of goods needed by consumers is equal to the quantities of goods provided by the sellers. This however does not hold in our case since the same data can be shared with more than one consumer at a time, which leads to an aggressive competition among data providers to sell their data even at lower prices\color{black}. The drawbacks of the merchant model are deply discussed in \citep{BATAINEH2019} which alternatively proposes a two sided market model for monetizing personal data. In more detail, Bataineh et al.\citep{BATAINEH2019} propose an open market model in which individuals (actual data owners) and data consumers trade data over a third party platform that helps them discover each other. The authors show that the two-sided market outperforms the merchant model in maximizing the total surplus. However, the main limitation of this approach is that it is based on a static analysis of consumer' demand and data prices, which makes it unsuitable for dynamic cloud markets\color{black}.

Under the umbrella of collaborative approaches, some proposals, for instance \citep{Chakareski2015} and \citep{Zehui2017}, tried to model the interactions among three entities in the domain of business-oriented IoT. In \citep{Chakareski2015}, the authors propose a model in which client peers are interested
in sharing video content with the help of the cloud. In \citep{Zehui2017}, the authors propose game theoretical models among IoT sensors, IoT service providers and data consumers. In these games, two entities (i.e., IoT sensors and IoT service providers) cooperate together in one game and then compete as one entity against data consumers. Such an approach suffers from three drawbacks: (1) it does not consider the cross-group externalities (e.g., the mutual impact of the clientele size of one party on that of the other party) among the involved parties, which makes it unable to capture the whole and more concrete and realistic picture of the three-sided economical model; (2) the cooperation and competition strategies adopted by the different players are highly impacted by the cross-group externalities which might not always lead to the best outcome for these players;  \color
{black}and (3) it does not clarify how cooperating entities would share their earned revenues.

Adopting traditional game theory concepts (e.g., Shapley value and Nash equilibrium) to distribute the revenue that results from the cooperation among the different parties suffers from several limitations when applied in dynamic data trading scenarios over the cloud. Specifically, 1) although such concepts might be highly efficient in scenarios wherein all the involved parties are rational, their effectiveness starts to decrease in the presence of parties that are heterogenious and prefer to deviate from the equilibrium points. For example, recent studies have revealed that only $37\%$ of the players tend to accept the Nash equilibrium in cooperative games (interested readers can consult  behavioral games and ultimatum games \citep{GUTH1982367} for further details); and 2) even though the Shapley value approach fairly splits the revenues among the cooperative entities based on their contributions, it becomes inapplicable in cases wherein the contributions of entities cannot be measured (which applies to the cloud scenario considered in this work). Specifically, the cloud provider adds an ethereal/intangible, yet significant, contribution to the coalition via introducing the wide social networks of data consumers to those of data providers. On other hand, data providers own the data which forms the core of this new business. This creates a continuous dilemma between data providers and cloud providers about who makes the most significant contribution to the coalition and hence who deserves the biggest share of the revenues. Equal distribution, so-called \emph{fifty-fifty}, is one approach to split the revenues between the cloud provider and data provider. However, as mentioned before, the rationality and greediness of the involved parties (i.e., the cloud provider and data provider) prohibit the success of such a strategy. This leads us to the conclusion that we are dealing with a behavioral and ultimatum game in which two players (proposer and responder) argue to split a certain amount of revenue. The proposer is endowed with a sum of revenue and is responsible for splitting this sum with the responder. The responder may accept or reject the sum. In the case the responder accepts the sum, the revenue is split as per the proposal; otherwise, both players receive nothing.



\color{black}



\begin{figure}[!h]

\includegraphics[scale = 0.45]{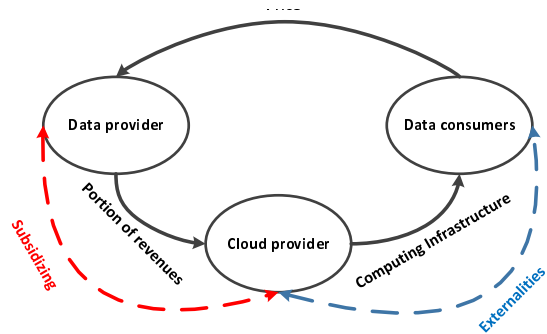}
    \caption{Overview of the proposed two-sided game}
    \label{game_overview}
\end{figure}

\textbf{Contributions.} To solve the aforementioned problems, the two-sided market model \citep{Rochet2003}, which is praised for its success in modeling situations that involve brokers and cross-group externalities, is investigated to study the cloud-based data trading problem. The main idea of our solution is that the cloud computing platform tries to attract data consumers by offering them higher amounts of computing resources to deploy their data analytic tasks. This in turn contributes in attracting a larger number of data providers to reach the cloud's network of data consumers. Consequently, the data providers have incentives to offer higher portions of their revenues to the cloud computing platform. Two-sided market provides effective solution concepts for situations that are characterized by a third-party platform connecting two other parties. However, the main limitation of the two-sided market theory is that it is effective in modeling scenarios in which the demand is static, but becomes less effective in elastic environments that characterize cloud computing where the demand is subject to dynamic and continuous changes. To address this problem, we integrate a novel game theoretical model, as shown in Figure \ref{game_overview}, on top of the two-sided market model. The players of our game are (multiple) independent competing service providers (followers) and the cloud computing platform (leader). The players opt for hybrid cooperative and non-cooperative strategies, where strategies are modeled as closed loops of dependencies. Data consumers and the cloud platform exhibit cross group externalities between each other, where a higher demand from consumers leads to a revenue increase for the cloud platform and a higher supply of computing resources from the cloud creates more demand from consumers.

In the first stage of the game, the leader (cloud platform) announces the desired portion of returned revenues out of the data providers' gain, and then in the second stage, data providers decide about their pricing strategy for data consumers. The resulting equilibrium forces the cloud platform to offer higher and reasonable supply of computing resources to guarantee maximal levels of revenues, while not showing greedy behavior in terms of its share of data providers’ revenue. Moreover, following our solution, the data providers are forced to offer the cloud platform a higher portion of their revenues to ensure appropriate Quality of Service (QoS) delivered to data consumers. In the case of a greedy behavior from the cloud, our game uses a subsidizing mechanism. This mechanism pushes data providers to increase the shared portion offered for the cloud to sustain high and reasonable levels of computing infrastructure so as to guarantee high levels of consumers' demand. Similarly, in cases where data providers behave greedily by offering small portions of revenues to the cloud, the subsidizing mechanism pushes the cloud to pump out more infrastructure units to increase the consumers' demand so as to guarantee the highest possible level of revenue portion.


To validate our solution, we conduct empirical experiments using real-world data from Google and Amazon. Experimental results show that by following our solution, all the involved parties (i.e., cloud platform, data providers and data consumers) achieve higher revenues than those achieved by the traditional cloud computing business model.

\section{Related Work}\label{Related_work}

In this section, we provide a literature review on cloud computing business models. The existing proposals can be classified into two main categories: classical market and game theoretic-based pricing models. The proposals under the classical market category such as \citep{Greenberg2008,Zhang2015,Rebai2015} tackle the pricing of the cloud services using simple pricing models including those of cost-based pricing, differential pricing, Ramsey pricing, and demand curve function. They model the pricing of cloud computing resources as an optimization problem among multiple cloud providers and cloud users. However, the main drawback of these approaches lies in their static pricing strategy which does not suit the highly variable and dynamic environment of cloud computing.

On the other hand, the game theoretical models consider the instantaneous interactions that might occur among the involved entities and their effects on each party's welfare. The objective is to dynamically capture the optimal price and distribution of the cloud computing resources. Many proposals such as \citep {Lin2015,Ranjan2013,Xu2012,Ding2015,Samimi2011} applied different approaches including games and machine learning \cite{wahab2021federated} to the cloud resource allocation and pricing problem. In \citep{Lin2015}, the authors propose an economic model based on a Stackelberg game to trade video contents and movies over a cloud platform. The proposed model formulates the interactions between a service provider (e.g., Netflix) and end users. The service provider acts as the game leader and aims to minimize the cloud bandwidth consumption while guaranteeing at the same time users' satisfaction. The work in \citep{Valerio2013} models the interactions among multiple Software as a Service (SaaS) providers and Infrastructure as a Service (IaaS) provider as a two-stage Stackelberg game. In the first stage of the game, SaaS providers determine the number of required VM instances while accounting for both the QoS delivered to their users and the associated costs. In the second stage, the IaaS providers seek to maximize their revenues in the light of the bids done by the SaaS providers \citep{Luong2017,Luong2016}. The author in \citep{Chakareski2015} proposes an economic model in which cloud users seek to share video content with other users over the cloud. The model is solved using both cooperative and non-cooperative games between the cloud and its users. Similar studies are investigated in \citep{Zehui2017,Ding2015} for different cloud applications. The authors in \citep{Niyato2020} propose a game theoretical model to deliver a bundle of complementary IoT services. The proposed solution studies the merchant-consumer scenario in which the IoT services are directly traded between the service providers and service consumers without the intervention from any third party. However, this solution cannot be adopted in our case, where the cloud computing is not the actual data owner and hence it cannot monetize the data directly for the consumers. Nevertheless, the cloud computing (the third party in our paper) is considered as a global market where the data services and data consumers meet each other, thus increasing their market shares. The authors in \citep{Bataineh2016} and \citep{Mashayekhy2014ATM} introduce a market model for managing, trading, and pricing big data services. Both proposals use the two-sided market theory in order to provide incentives for both cloud providers and users to increase their data shares. The work presented in \citep{BATAINEH2019} extends the work proposed in \citep{Bataineh2016} and comprehensively studies the two-sided market model as a successful model for monetizing personal data. However, these proposals consider a static environment in which the demands on cloud resources are computed in a static manner, which makes them unable to accommodate the cloud's elasticity property.

To the best of our knowledge, the proposed work is the first that addresses big data services monetization, while considering the cross-group externalities among the involved entities. Unlike the classical cloud computing business model (where the main challenge is how to optimize the cloud utilization while incorporating only operational cost and QoS metrics), our approach : 1) supports and helps junior big data service providers especially those that have limited monetary budgets; 2) uses the two-sided market theory to model the interactions among the involved parties, while all above-discussed proposals use the classical merchant model; (3) includes a subsidizing technique to push the resulting equilibrium toward a Pareto optimal point. On the other hand, the above-discussed proposals adopt the fairness criterion that rewards the involved parties based on their contributions. We also differ from the other proposals that adopt the two-sided market theory by providing a dynamic pricing method, instead of  a static game theoretic-based pricing strategy.

\section{Proposed Big Data Services Monetization Model over the Cloud: A Two-sided Game Model}\color{black}
We explain in this section the details of our proposed Big data services monetization.
\subsection{Solution Architecture and Game Formulation}\label{TM-settings}

\begin{figure}[h]%

\includegraphics[scale = 0.5]{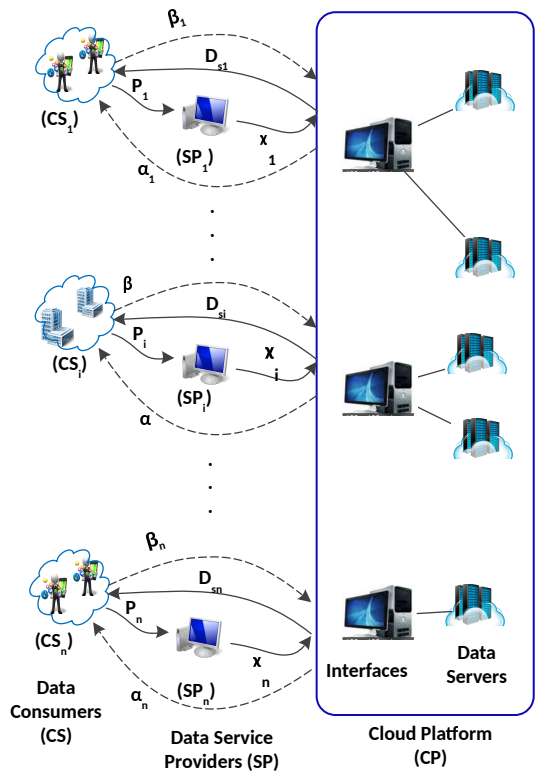}
\caption {Two-sided model}
\label {Two-sided market platform}
\end{figure}%



The proposed cloud market platform, depicted in Figure \ref{Two-sided market platform}, consists of three entities:  consumers of services $CS$ ($CS_i$ denotes Consumers of Service $i$), big data service providers $SP$ (a Service Provider providing service $i$ is denoted $SP_i$) and a typical Cloud Platform ($CP$). The cloud platform, such as Google and Amazon, is a market leader with huge computing and storage capabilities, capitals, and social consumer networks. In our model, a certain big data service provider $SP_i$ that provides a service $i$ deploys its service on the cloud and receives a monetary value of $P_i$ for each  consumer access to its service $i$. The cloud platform $CP$ is in charge of sustaining the consumer access through providing the needed computing and storage infrastructure including hardware, software and security services. The relationship between consumer $CS_i$'s demand, denoted by $D_{c_i}$, and the computing and storage resources $D_{s_i}$ supplied to $CS_i$ is modeled using the two-sided market model as cross group externalities $\alpha$ and $\beta$. Here, $\alpha$ represents the benefits that a consumer obtains when some new computing and storage resources are added to $D_{s_i}$ and $\beta$ represents the amount of benefits that the cloud platform earns when more new consumers are added to $D_{c_i}$. The parameters $\alpha$ and $\beta$ are dependant on the service $i$. However, instead of using the notations $\alpha_i$ and $\beta_i$, the index $i$ is omitted to simplify the equations where the service $i$ is understood from the context. The same simplification is used for the other parameters that appear as powers (exponents) in our equations.

The interaction between $SP$ and $CP$ is modeled as a two-stage game where $CP$ acts as the game leader and $SP$ are the followers. In the first stage of the game, each service provider providing service $i$ $SP_i$ observes the amount of money returns $\chi_i$ requested by $CP$, in order to adjust the price to be charged to $CS_i$. In quest of the price specified by $SP_i$, $CP$ determines the optimal amount of computing and storage resources $D_{s_i}$ that should be supplied to $CS_i$. The model forms a closed loop of dependencies that involves techniques from Stackelberg and Ultimatum game theory as well as a subsidizing technique.

In the Stackelberg game, the interactions take place in two stages where the leader ($CP$) makes the first move and then does each follower ($SP_i$) after having observed the leader's move. In the ultimatum game, the first player ($CP$) proposes a strategy to divide the amount of returned revenue with the second player ($SP_i$). In case $SP_i$ rejects the offer, neither player gains anything. Otherwise, the first player gets the amount it requested and the second player gets the rest. In the subsidizing technique,  $SP_i$ may chose to subsidize $CP$ by an extra amount of payment that exceeds the contribution of this $CP$. The objective is to keep an optimal level of $D_{s_i}$ that maximizes the return revenues $P_i * D_{c_i}$. Alternatively, $CP$ may subsidize $SP_i$ by low portion of the resulting revenues to keep an optimal level of $P_i$. The different parameters and symbols used in our proposed solution are depicted in Table \ref{tab:ModelParameters}.

\begin{table}[h]
	\centering
		\begin{tabular}{|c|c|}
		\hline
		\hline
			\scriptsize  Model Parameters
				&\scriptsize Descriptions. \\
				\hline
				\hline
			\scriptsize  $ SP_{i}$
				&\scriptsize Service provider providing service $i$.

              \\  \scriptsize  $ CP$
				&\scriptsize A typical cloud platform.

\\  \scriptsize $CS_{i}$
        & \scriptsize  Consumers of service $i$.

                \\ \scriptsize  $D_{c_i}$
				&\scriptsize $CS_i$'s demand.
				
				\\  \scriptsize $D_{s_i}$
        & \scriptsize  IT-infrastructure supply to $CS_i$.

        \\ \scriptsize  $ P_i $
				&\scriptsize Service $i$'s price.

        \\ \scriptsize $\phi$
        & \scriptsize $D_{s_i}$'s elasticity with respect to $\chi_{i}$.

		\\  \scriptsize $\chi_i$
        & \scriptsize Portion of revenue required by $CP$ from $SP_{i}$.

        \\
				
					\scriptsize  $\gamma$
				&\scriptsize $D_{c_i}$'s elasticity with respect to $P_{i}$.
				 \\ \scriptsize $\beta$
        & \scriptsize The Network effects (externality) on $D_{s_i}$ by $D_{c_i}$

				\\

								\scriptsize  $\psi$
				&\scriptsize $D_{s_i}$'s elasticity with respect to $P_{i}$.
				\\ \scriptsize  $\alpha$
				&\scriptsize The Network effects (externality) on $D_{c_i}$ by $D_{s_i}$.

        \\
				
				\scriptsize  $f_{c}$
				&\scriptsize Associated costs per service consumer.

           \\       \scriptsize $k_1$, $k_2$
        & \scriptsize Constant multipliers.

        \\
				
				\scriptsize  $\pi_i$
				&\scriptsize $SP_i$'s payoff.

                \\  \scriptsize $f_s$
        & \scriptsize Associated costs per IT-infrastructure unit.

	%
				
                \\  \scriptsize $\pi$
        & \scriptsize Cloud platform's payoff.

\\ \scriptsize    $a_1$
				& \scriptsize $= \gamma - \alpha \psi$.

        \\  \scriptsize $a_2$
        & \scriptsize $= 1 - \alpha \beta$.

               \\

\scriptsize    $a_3$
				&\scriptsize $ = \psi - \gamma \beta$.

        \\

\scriptsize $a_4$
                & \scriptsize  $= \alpha \phi$
 \\


				\hline
		\end{tabular}
	\caption{Model parameters }
	\label{tab:ModelParameters}
\end{table}

\subsection{Players' Demand and Utility Functions}

A precise estimation of the needed  computing and storage resources requires a price estimation mechanism for the number of consumers and the variation of their demand with respect to the provided QoS. To do so, we define the consumer's demand and supply using the Cobb-Douglas function that effectively captures the elasticity of the computing and storage resources supply ($D_{s_i}$) and its variations for each specific user's demand. This elasticity is a characteristic property of cloud computing environments. The demand functions we use are continuous, concave or convex, and  capture the elasticity with respect to each input parameter. Two elasticity parameters are used $\gamma$ and $\psi$ (see Table \ref{tab:ModelParameters}). These two parameters depend on the service $i$, which is omitted from the notations for simplicity as mentioned earlier. In our model, the consumer's demand ($D_{c_i}$) is a function of $P_i$ and $D_{s_i}$ as shown in Equation (\ref{consumer_demand}).


\begin{equation} \label{consumer_demand}
D_{c_i} = k_1  P_{i}^{- \gamma}  D_{s_i}^{\alpha}
\end{equation}

$D_{s_i}$ is given in Equation (\ref{Supply_demand}). Clearly, higher consumers' demands would have an influence on the quantity of supplied resources. The cloud platform $CP$ uses more computing and storage resources to keep up with the increasing number of consumer accesses, to maintain a high quality level. The parameter $\chi_{i}^{\phi}$ represents the cloud platform's preferences (i.e., desired profit) and implicitly captures the rationality of both $CP$ and $SP_i$. In fact, it reflects the level of perfect/imperfect information that $CP$ and $SP_i$ have about one another. High elasticity $\phi$ is caused either by a greedy monopolist cloud platform or by a weak service with few capitals accepting small portions of returns on profits. The parameter $\phi$ depends on the service $i$, but as mentioned earlier, the index $i$ is omitted when the service $i$ is understood from the context. The charged price $P_i$ also positively contributes to $D_{s_i}$. We can arguably claim that charging consumers with higher prices $P_i$ forces $CP$ to provide more computing and storage resources so as to satisfy the consumers' needs. Modeling $D_{s_i}$ as a function of $\chi_i$ and $P_{i}$ with different elasticity values connects $CP$ and $SP_i$ strategies with each other, which captures the sensitivity of $CP$ to $SP_i$'s strategy (i.e., structure of the charged price and shared portion), and highlights the importance of the subsidizing technique. This aspect is illustrated and discussed further in the simulation section (Section \ref{sensivity_anylsis_subsidizing_factor}).


\begin{equation} \label{Supply_demand}
D_{s_i} = k_2  \chi_{i}^{\phi}  P_{i}^{\psi}  D_{c_i}^{\beta}
\end{equation}

\noindent By substituting Equation (\ref{Supply_demand}) into Equation (\ref{consumer_demand}) and vice versa, we can express $D_{c_i}$ and $D_{s_i}$ as functions of $P_i$ and $\chi_i$ as follows:

\begin{equation}\label{consumer_demand1}
D_{c_i} = (k_1  k_2^{\alpha}  P_{i}^{-a_1}  \chi_{i}^{a_4} )^{1/a_2}
\end{equation}

\begin{equation}\label{supply_demand2}
D_{s_i} = (k_2  k_1^{\beta} P_{i}^{a_3}  \chi_{i}^{\phi})^{1/a_2}
\end{equation}

Each big data service provider $SP_i$ is subject to a fixed cost $f_c$ per each consumer access. $SP_i$ aims to maximize its payoff as given in Equation (\ref{service_payoff}). We express the service provider's payoff $\pi_{i}$ as a function of $P_i$ and $\chi_i$ by substituting Equation (\ref{consumer_demand1}) into Equation (\ref{service_payoff}) and taking the $\log$ for both sides as shown in Equation (\ref{service_payoff1}).

\begin{equation} \label{service_payoff}
 \pi_{i} = ((P_i)(1-\chi_i)- f_c) D_{c_i}
\end{equation}

\begin{equation} \label{service_payoff1}
\begin{multlined}
\log \pi_{i} = \log (P_i (1-\chi_i) - f_c) + (1/a_2)(\log k_1 k_2^{\alpha}
\\ - a_1 \log P_i + a_4 \log \chi_i)
\end{multlined}
\end{equation}

The cloud platform $CP$ is subject to a fixed cost $f_s$ per each unit of computing and storage resources. The $CP$ aims to maximize its payoff as given in Equation (\ref{cloud_payoff}). We express the cloud platform's payoff $\pi$ as a function of $P_i$ and $\chi_i$ by substituting Equations (\ref{consumer_demand1}) and (\ref{supply_demand2}) into Equation (\ref{cloud_payoff}) as shown in Equation (\ref{cloud_payoff1}).

\begin{equation}\label{cloud_payoff}
\pi= P_{i} \chi_{i}  D_{c_i} - f_s  D_{s_i}
\end{equation}

\begin{equation}\label{cloud_payoff1}
\pi = (k_1 k_2^{\alpha})^{\frac{1}{a_2}}  P_{i}^{1- \frac{a_1}{a_2}}  \chi_{i}^{\frac{a_4}{a_2}+1} - f_s ((k_2  k_1^{\beta})^{\frac{1}{a_2}}  P_i^{\frac{a_3}{a_2}}  \chi_i^{\frac{\phi}{a_2}} )
\end{equation}

\subsection{Game Equilibrium}

The equilibrium of the above-described game is solved using a backward induction methodology. Thus, the followers' (service providers) sub-game is solved first to obtain their response $P_i$ to the service consumers. The leader's (cloud platform) sub-game is then computed considering all the possible reactions of its followers to maximize its payoff \citep{wahab2016stackelberg}. Every service provider $SP_i$ determines its optimal decision $P_{i}^{*}$, while considering the $CP$'s optimal decision $\chi_{i}^{*}$ as an input parameter. The players' best responses are discussed in the following.

\begin{theorem}\label{game_equ}
The best responses in the two-sided game are as follows:
\begin{enumerate}
\item The best response of the service provider $SP_i$ is given by:
\begin{equation}\label{service_best_response}
P_i^{*} = \frac{a_1 f_c}{ (a_1 - a_2)(1 - \chi_i)}
\end{equation}

if: ~~ $\frac{a_1}{a_1 - a_2} > 0$ ~and~ $\frac{a_1}{a_2} > 1$

\vspace{0.5cm}

\item The best response of the cloud platform with respect to a service $i$ is given by:
\begin{equation}\label{cloud_best_response}
\begin{multlined}
\chi_{i}^{\frac{a_4 - \phi}{a_2} + 1 } (1-\chi_i)^{\frac{a_1 + a_3}{a_2} - 1} = f_s \times (\frac{\phi}{a_4 + a_2})
\\
\times (\frac{k_2 k^{\beta}_1}{k_1 k^{\alpha}_2} )^{\frac{1}{a_2}}
 \times (\frac{a_1 f_c}{a_1 - a_2})^{\frac{a_1 + a_3}{a_2} -1}
\end{multlined}
\end{equation}

if: ~~ $a_4 + a_2 - \phi < 0$

\end{enumerate}

\begin{proof}
Consider the service payoff given by Equation (\ref{service_payoff1}), the optimal price $P_{i}^{*}$ is defined by $\partial \pi_{i} / \partial P_i  = 0$ as follows:









\begin{equation}\label{optimalservice}
\frac{1}{\pi_i} \times \frac {\partial{\pi_i}}{\partial{P_i}} = \frac{1-\chi_i}{P_i(1 - \chi_i) - f_c}
- \frac{a_1}{(a_2) P_i} = 0
\end{equation}

$\Rightarrow $
\begin{equation}\label{optimalprice}
P_i^{*} = \frac{a_1 f_c}{ (a_1 - a_2)(1 - \chi_i)}
\end{equation}

\noindent Since $P_i^{*}$ is always positive, then
\begin{equation}
\frac{a_1}{a_1 - a_2} > 0
\end{equation}

\noindent To verify the type of $P_i^{*}$'s optimality, i.e maximum or minimum, we compute a second derivative test by deriving Equation (\ref{service_payoff}):

\begin{equation}\label{derive_servicepayoff_orginal}
    \frac{\partial \pi_{i}}{\partial P_i} = (1- \chi_{i}) D_{c_{i}} + (P_i(1 - \chi_i) - f_c) \frac{\partial D_{c_{i}}}{\partial P_i}
\end{equation}
By deriving Equation \ref{consumer_demand1}, then

\begin{equation}\label{der_consumer_demand1_new}
    \frac{\partial D_{c_{i}}}{\partial P_i} = \frac{-a_{1}}{a_{2} P_i} D_{c_{i}}
\end{equation}
By substituting Equation \ref{der_consumer_demand1_new} into Equation \ref{derive_servicepayoff_orginal}, then

\begin{equation}\label{first_der}
\frac{\partial \pi_{i}}{\partial P_i} = (1- \chi_{i}) D_{c_{i}} - \frac {a_1}{a_2 P_i} (P_i(1 - \chi_i) - f_c) D_{c_i}
\end{equation}

\noindent By rewriting Equation (\ref{first_der}) using Equation (\ref{service_payoff}), then

\begin{equation}\label{first_der1}
\frac{\partial \pi_i}{\partial P_i} = (1- \chi_{i}) D_{c_i} - \frac {a_1}{a_2 P_i} \pi_{i}
\end{equation}

\begin{equation}\label{second_der}
\frac{\partial^2 \pi_i}{\partial P^{2}_i} = \frac{ - (1-\chi_i) a_1}{a_2 P_{i}} D_{c_{i}} - \frac{a_1}{a_2 P_{i}} \frac{\partial \pi_{i}}{\partial P_i} + \frac{a_1}{a_2 P^2_{i}} \pi_i
\end{equation}

\noindent By simplifying Equation (\ref{second_der}) and substituting Equation (\ref{optimalprice}), we obtain:
\begin{equation}\label{second_der1}
\frac{\partial^2 \pi_i}{\partial P^{2}_i} = \frac{D_{c_i}}{P_i} (1-\frac{a_1}{a_2}) (1-\chi_{i})
\end{equation}

\noindent Since $D_{c_i}$ and $P_i$ are always positives, then

\begin{equation}\label{first_condition}
\frac{\partial^2 \pi_i}{\partial P^{2}_i} < 0 \Rightarrow  (1-\frac{a_1}{a_2}) < 0  \Rightarrow  \frac{a_1}{a_2} > 1
\end{equation}

\noindent Similarly, to obtain the optimal $\chi^{*}_{i}$, we derive Equation (\ref{cloud_payoff1}) with respect to $\chi_i$ as given by Equation (\ref{cloud_first_der}): 

\begin{equation}\label{cloud_first_der}
\frac{\partial \pi}{\partial \chi_i} = (k_1 k_2^{\alpha})^{\frac{1}{a_2}}(\frac{a_4}{a_2} + 1) P_i^{1-\frac{a_1}{a_2}} \chi^{\frac{a_4}{a_2}}_{i} - (\frac{\phi f_s (k_2 k_1^{\beta})^{\frac{1}{a_2}}}{a_2} ) P_i^{\frac{a_3}{a_2}} \chi^{\frac{\phi}{a_2} - 1}_{i} = 0
\end{equation}

\begin{equation}\label{optimal_cloud}
\chi_{i}^{\frac{a_4 - \phi}{a_2} + 1} =  f_s (\frac{\phi}{a_4 + a_2}) (\frac{k_2 k^{\beta}_1}{k_1 k^{\alpha}_2})^{\frac{1}{a_2}} P^{\frac{a_1 + a_3}{a_2} -1}_{i}
\end{equation}

\noindent By substituting Equation (\ref{optimalprice}) in Equation (\ref{optimal_cloud}), we get:

\begin{equation}\label{optimal_cloud1}
\chi_{i}^{\frac{a_4 - \phi}{a_2} + 1 } (1-\chi_i)^{\frac{a_1 + a_3}{a_2} - 1} = f_s (\frac{\phi}{a_4 + a_2}) (\frac{k_2 k^{\beta}_1}{k_1 k^{\alpha}_2} )^{\frac{1}{a_2}} (\frac{a_1 f_c}{a_1 - a_2})^{\frac{a_1 + a_3}{a_2} -1}
\end{equation}

\noindent To verify the type of $\chi_{i}^{*}$'s optimality, we compute a second derivative test by deriving Equation (\ref{cloud_first_der}) as given by Equation (\ref{cloud_second_der}):

\begin{equation}\label{cloud_second_der}
\begin{multlined}
\frac{\partial^{2} \pi}{\partial \chi_{i}^{2}} = (k_1 k_2^{\alpha})^{\frac{1}{a_2}} (\frac{a_4 (a_4 + a_2) }{a^{2}_2}) P^{\frac{a_2 - a_1}{a_2}}_{i} \chi^{\frac{a_4}{a_2} - 1}_{i}
\\ - (\frac{ f_s \phi (\phi - a_2) (k_2 k_1^{\beta})^{\frac{1}{a_2}}} {a^2_2} ) P_i^{\frac{a_3}{a_2}} \chi^{\frac{\phi}{a_2} - 2}_{i} < 0
\end{multlined}
\end{equation}

\noindent By substituting Equation (\ref{optimal_cloud}) in Equation (\ref{cloud_second_der}), we obtain:

\begin{equation}\label{condition_3}
a_4 + a_2 - \phi < 0
\end{equation}

\end{proof}

\end{theorem}

\section{Simulations and Empirical Analysis}

\begin{figure}[!h]

    \includegraphics[scale = 0.44]{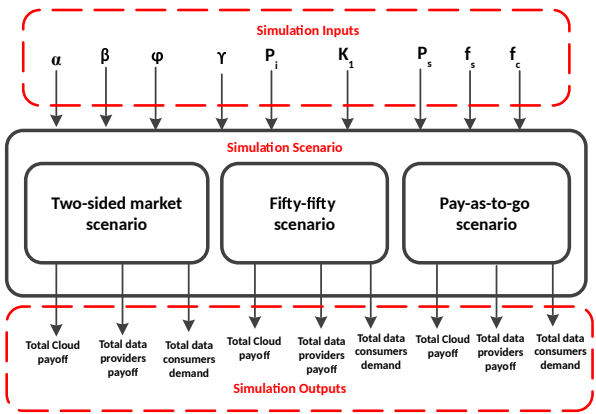}
    \caption{Simulation overview}
    \label{Simulation_overview}
\end{figure}

In this section, we evaluate the performance of the proposed two-sided game solution in comparison with the fifty-fifty and the pay-as-to-go approaches in terms of total surpluses of involved parties (i.e., payoffs of the cloud platform, service data providers and service data consumers). Specifically, we aim to: 1) verify the effectiveness of the proposed game vis-à-vis the current cloud computing business model (i.e., the pay-as-to-go model);  2) study the equilibrium of the two-sided game in presence of the fifty-fifty choice (i.e., egalitarian choice in ultimatum games) which is the typical solution for such games; and 3) investigate the impact of the model parameters on the performance of our solution. Figure \ref{Simulation_overview} shows an overview of our simulation setting in terms of inputs, scenarios, and results. The simulation inputs and scenarios are described in Sections \ref{Simulation_Setup} and \ref{Simulation_Scenarios} respectively,  while the simulation results are discussed in detail in Sections \ref{Sensitivity_analysis_of_externalities}, \ref{sensivity_anylsis_subsidizing_factor} and \ref{sensivity_anylsis_gamma}. 




\subsection{Simulation Setup}\label{Simulation_Setup}

In this section, we conduct a simulation analysis grounded on statistical observations from BMR \citep{DMRAMAZON2018} and real data from \citep{google2019}. According to \citep{DMRAMAZON2018}, in $2019$, Amazon Web services (AWS) received $30$ billion USD in revenue with net income around $10$ billion USD from $1$ million active customers running monthly $70$ million hours of their enterprises on custom instances of Elastic Compute Cloud (EC2). So, 1) an enterprise customer spends on average up to $30,000$ USD per year in monthly renting $70$ hours of cloud resources; and 2) the marginal operating costs for the cloud platform is $66\%$ of revenues (amazon received $10,000$  USD as net payoff from each consumer)\color{black}. By entering these numbers in the Amazon calculator \citep{Amazon_Calculator}, we can conclude that the customer rents on average $70$ hours monthly of $32$ instances of Amazon EC2 where each instance includes $16$ VMs, $30$ GB of Memory, and $1000$ GB of hard disk storage at rate $36$ USD$/$hour\color{black}. The price rate ($36$ USD$/$hour) is denoted by $P_s$, which will be used later to calculate the cloud and data providers payoffs in the pay-as-to-go model as explained in Section \ref{Simulation_Scenarios}\color{black}. The cloud provider (amazon) entails $66\%$ of instances price ($36$ USD$/$hour) as operating costs, which is $23.7$ USD$/$hour. The operating costs are denoted in our model by $f_s$\color{black}. In fact, $40\%$ of revenues as a profit and $60\%$ as an operation cost are common in business. Thus, we assume the marginal cost of data consumers ($f_c$) entailed by data providers has the same distribution as ($f_s$). The enterprise customer and its consumption of EC$2$ instances are represented by $SP_i$ (service data provider) and $D_{s_i}$ respectively. The mean of the supply function $D_{s_i}$ consists of $32$ EC$2$ instances. However, enterprise customers have varying business types and hence vary in terms of the amount of needed cloud resources. To model this variation in our simulations, the customers' demand on EC$2$ instances is normally distributed around the mean with a standard deviation of $10$. This means that the co-domain of the supply function $D_{s_i}$ ranges from $1$ to $53$ EC$2$ instances. The real dataset \citep{google2019} registers the log file of computational big data jobs executed by tremendous enterprise customers over similar instances of EC2. This dataset helps us extract reliable ranges of consumers' demands $D_{c_i}$ as well as the externalities $\alpha$ and $\beta$ as described in what follows. The computational power of each instance, extracted from the same dataset \citep{google2019}, is normally distributed with a mean of $0.38$ job per second and a standard deviation of $0.1$. The average computational power is represented in the proposed model by the externality factor $\alpha$, which means that $\alpha$ ranges from $0.1$ to $0.7$. According to the assumption presented in \citep{Rochet2003}, the cross group externalities factor should be bounded by $0$ and $1$, i.e., $ 0<\alpha \beta < 1$. Hence, the externality factor $\beta$ would range from $0$ to $1/\alpha$. A consumer's demand $D_{c_i}$ on each enterprise ranges from $0.1 \times 1$ to $0.7 \times 53$, which is $0.1$ to $37$ requests per second. The service price, $P_i$, is estimated through observing the prices of $150$ business intelligence computing services including big data and IoT services located in the the AWS marketplace \citep{Amazon_marketplace}. According to the observed prices, $P_{i}$ is normally distributed with a mean of $1.7$ USD$/$hour and a standard deviation of $0.5$ USD. This means that the service prices range from $0.2$ USD$/$hour to $3.2$ USD$/$hour. The parameter $\phi$ represents the greediness of the cloud platform with respect to the service providers. The subsidizing factor $0<\phi<1$ represents the rational behavior (subsidizing behavior) of the cloud, while $1<\phi$ represents the greedy behavior of the cloud platform\color{black}. The price elasticities are set up between $0.1 - 0.35$ (i.e., $\gamma$), which are similar to the sensitivity of mobile/telecommunication services price shown in the literature \citep{Danaher2002}. We assume that $k_2 = 1$ in our simulation. By substituting the expected values of $\alpha$, $D_{s_i}$, $D_{c_i}$, $\gamma$, and $P_i$ into Equation (\ref{consumer_demand}) and considering the assumption $(k_2 = 1)$, we find that the multiplier $k_1$ ranges from $0.1$ to $0.99$. It is worth mentioning that consumers demand $(D_{c_i})$ and cloud resources (i.e., computing and storage resources) $D_{s_i}$ supplied to $CS_i$ are only estimated under simulation setup to extract a suitable range for the multiplier $k_1$, but they are not given as simulation inputs. The simulation calculates the expected consumer demand and the optimal supply of cloud resources as explained in Section \ref{Simulation_Scenarios}. The values of all associated parameters are summarized in Table \ref{tab:system_parameters_simulation}.

\color{black}
\begin{table}[!h]
\begin{center}
\begin{tabular}{ |c|c| }
 \hline
\hline
System Parameters & Values  \\
 \hline
\hline
 $P_{i}$ & $0.2$ - $3.2$ USD per hour  \\
$\phi$ & $0$ - $5$  \\
$\alpha$ & $0.1$ - $0.7$  \\
$\beta$ & $0$ - $1/\alpha$  \\
$\gamma$ & $0$ - $0.35$  \\
$P_s$ & 36 USD per hour \\
$k_1$ & $0.1 - 0.9$ \\

 \hline
\hline
\end{tabular}
\end{center}
\caption{Simulation parameters values}
\label{tab:system_parameters_simulation}
\end{table}

\subsection{Simulation Scenarios} \label{Simulation_Scenarios}
\color{black} We consider a group of $300$ data service providers in the cloud under three scenarios: 1) proposed two-sided game; 2) fifty-fifty scenario which follows our model except that the cloud platform and data provider agree to share the revenue equality; and 3) pay-as-to-go scenario, which is the current business model adopted by the main cloud providers such as Amazon and Google.\color{black}

\subsubsection{Two-sided scenario}

The two-sided model, explained in details in Section \ref{TM-settings}, is described in Algorithm \ref{twosidedscenario}. Given a data service price $P_i$, the cloud platform determines the optimal portion of revenue $\chi_i$ and the required amount of cloud resources that maximize its payoff.

\begin{algorithm}
	\caption{Two-sided scenario}
	\label{twosidedscenario}
	\textbf{Input:} $\alpha$, $\beta$, $\phi$,$\gamma$, $k_1$, $f_s$, $f_c$ \\
	\textbf{Output:} \textit{CloudPayoff}, \textit{ProvidersPayoff}, \textit{ConsumersDemand}
	\begin{algorithmic}[1]
	
		\ForEach {$Data \, service \, provider \, SP_i$}
			\State $SP_i$ declares its price $P_i$
			
			\State $\chi_i$ is calculated through maximizing Equation \eqref{service_payoff}
			\State The cloud calculates $D_{s_i}$  through maximizing Equation \eqref{cloud_payoff}
			\State Equation \eqref{consumer_demand} is used to determine \textit{ConsumersDemand}
			\State Equation \eqref{service_payoff} is used to determine \textit{ProvidersPayoff}
			\State Equation \eqref{cloud_payoff} is used to determine \textit{CloudPayoff}
			
		\EndFor
	\end{algorithmic}
\end{algorithm}

\subsubsection{Fifty-fifty scenario}
The egalitarian (fifty-fifty) scenario follows the two-sided market model in terms of consumer demand and supply function, thus considering the externalities among the involved parties (i.e., Equations \eqref{consumer_demand} and \eqref{Supply_demand}). Thus, the utilities of the cloud platform and data providers are formalized using the same payoff equations used in the two-sided model (i.e., Equations \eqref{cloud_payoff} and \eqref{service_payoff} ). However, under this scenario, the cloud platform requests $50\%$ of the revenue and hence the subsidizing factor $\phi$ is reset by $1$. The fifty-fifty scenario is described in Algorithm \ref{fiftyscenario}. Specifically, given a shared portion $\chi_i = 0.5$ and a subsidizing factor $\phi=1$, the data provider determines its optimal service price $P_i$ through maximizing its payoff given in Equation \eqref{service_payoff}. Thereafter, the cloud platform calculates the optimal computing infrastructure $D_{s_i}$ by maximizing its payoff given in Equation \eqref{cloud_payoff}.

\begin{algorithm}
	\caption{Fifty-fifty scenario}
	\label{fiftyscenario}
	\textbf{Input:} $\alpha$, $\beta$,$\gamma$, $k_1$, $f_s$, $f_c$ \\
	\textbf{Output:} \textit{CloudPayoff}, \textit{ProvidersPayoff}, \textit{ConsumersDemand}
	\begin{algorithmic}[1]
	
		\ForEach {$Data \, service \, provider \, SP_i$}
			
			\State $ \chi_i \gets 0.5$
			\State $ \phi \gets 1$
			\State  $SP_i$ declares $P_i$ through maximizing Equation \eqref{service_payoff}
			\State The cloud calculates $D_{s_i}$  through maximizing Equation \eqref{cloud_payoff}
			\State Equation \eqref{consumer_demand} is used to determine \textit{ConsumersDemand}
			\State Equation \eqref{service_payoff} is used to determine \textit{ProvidersPayoff}
			\State Equation \eqref{cloud_payoff} is used to determine \textit{CloudPayoff}
			
		\EndFor
	\end{algorithmic}
\end{algorithm}

\subsubsection{Pay-as-to-go model}
\begin{figure}[!h]
    \includegraphics[scale = 0.45]{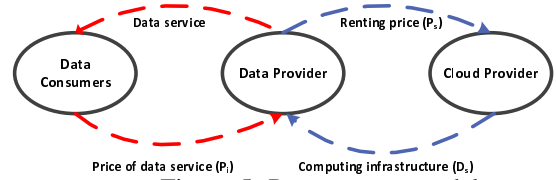}
    \vspace{-2.5cm}
    \caption{Pay-as-to-go model}
    \label{pay_as_to_go_model}
\end{figure}

Figure \ref{pay_as_to_go_model} depicts the pay-as-to-go model. As shown in the figure, the data provider rents the cloud computing infrastructure $(D_{s_i})$ for a price of $P_s$ USD$/$hour. Thereafter, the data provider delivers its own data services to the data consumers for a price of $P_i$ USD$/$hour.
The data provider and cloud payoffs under the pay-as-to-go model are given in Equations \eqref{provider_payoff_under_payastogo} and \eqref{cloudpayoff_underpayastogo} respectively. The consumer's demand function under the pay-as-to-go model has the same characteristics as under the two-sided market model. This means that the consumer's demand is formalized as given in Equation \eqref{consumer_demand}. However, the provided cloud computing infrastructure $(D_{s_i})$ is not given as a function under the pay-as-to-go model since the cloud platform and data consumers do not directly interact nor do they exhibit mutual cross group externalities between each other. The pay-as-to-go model is described in Algorithm \ref{payastogoalg}. Given a price rate of $P_s$ USD$/$hour, the data provider determines the optimal amount of rented cloud computing infrastructure through maximizing Equation \eqref{provider_payoff_under_payastogo}. The optimal amount of rented cloud computing infrastructure $D^{*}_{s_i}$ is given in Equation \eqref{optimal_ds} through substituting Equation \eqref{consumer_demand} with Equation \eqref{provider_payoff_under_payastogo} and computing its derivatives with respect to $D_{s_i}$.

\color{black}

\begin{equation}\label{provider_payoff_under_payastogo}
   \pi_{i} = (P_i - f_c) D_{c_i} - P_s D_{s_i}
\end{equation}

\begin{equation}\label{cloudpayoff_underpayastogo}
\pi= (P_s - f_s )D_{s_i}
\end{equation}

\begin{equation}\label{optimal_ds}
    D^{*}_{s_i} = (\frac{P_s}{\alpha k_1 (P_i - fc) P^{-\gamma}_i} )^{\frac{1}{\alpha -1}}
\end{equation}

\begin{algorithm}
	\caption{Pay-as-to-go scenario}
	\label{payastogoalg}
	\textbf{Input:} $\alpha$, $\beta$,$\gamma$, $P_s$, $k_1$, $f_s$, $f_c$ \\
	\textbf{Output:} \textit{CloudPayoff}, \textit{ProvidersPayoff}, \textit{ConsumersDemand}
	\begin{algorithmic}[1]
		\ForEach {$Data \, service \, provider \, SP_i$}
			\State The cloud declares $P_{s}$
			\State $SP_i$ declares $P_i$
			\State $SP_i$ calculates $D_{s_i}$ using Equation \eqref{optimal_ds}
			\State Equation \eqref{consumer_demand} is used to determine \textit{ConsumersDemand}
			\State Equation \eqref{provider_payoff_under_payastogo} is used to determine \textit{ProvidersPayoff}
			\State Equation \eqref{cloudpayoff_underpayastogo} is used to determine \textit{CloudPayoff}
			
		\EndFor
	\end{algorithmic}
\end{algorithm}

\subsection{Sensitivity analysis of externalities} \label{Sensitivity_analysis_of_externalities}
\begin{figure}[!h]
    \includegraphics[scale = 0.5]{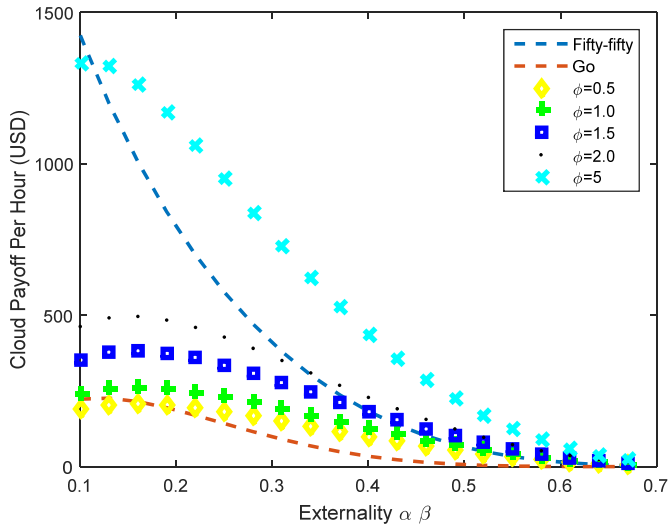}
    \caption{Cloud payoff over externalities $\alpha \beta$}
    \label{clod_payoff_externalities}
\end{figure}

\begin{figure}[!h]
    \centering
    \includegraphics[scale = 0.5]{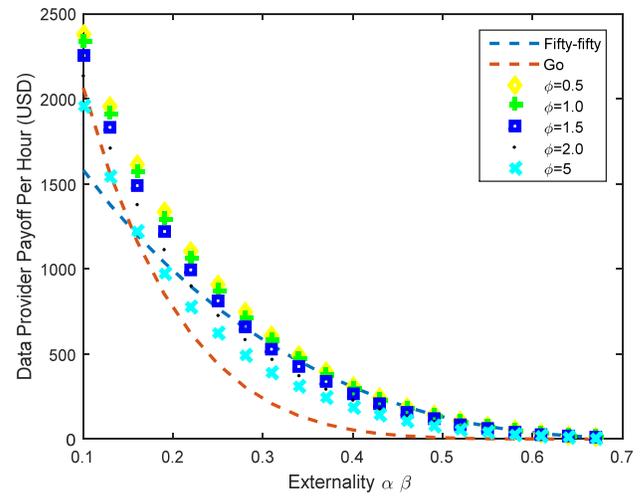}
    \caption{Data providers payoff over externalities $\alpha \beta$}
    \label{data_provider_payoff_externalities}
\end{figure}

\begin{figure}[!h]
    \centering
    \includegraphics[scale = 0.5]{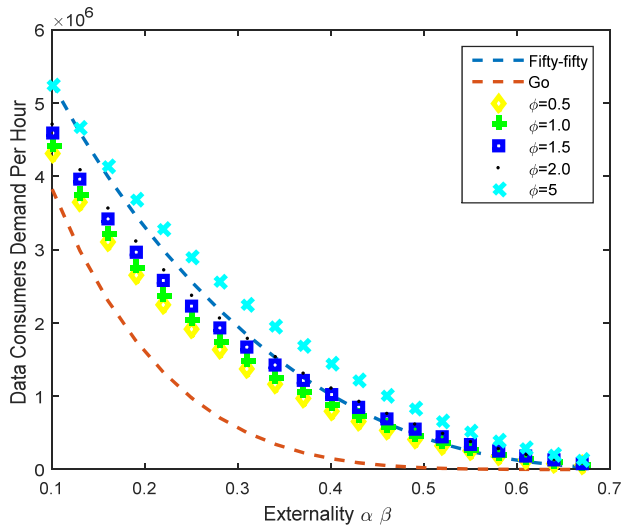}
    \caption{Data consumers demand over externalities $\alpha \beta$}
    \label{data_consumer_demand_over_externalities}
\end{figure}

We first investigate in \color{black}Figures \ref{clod_payoff_externalities}, \ref{data_provider_payoff_externalities}, and \ref{data_consumer_demand_over_externalities} \color{black}the impact of the externality parameters $(\alpha \beta)$ on the payoffs of the cloud platform, data providers, and data consumers respectively. We run the simulation with different ranges of the subsidizing factor $(\phi)$, i.e., \color{black} $0.5$, $1.0$, $1.5$, $2.0$, and $5.0$\color{black}. Those values of externality and subsidizing factor are given as inputs to the simulation program to adjust the strategies of the involved players. Specifically, the cloud platform adjusts the amount of provided infrastructure $(D_{s_i})$ and the demanded portion $(\chi_{i})$, while the data provider $(SP_i)$ calculates the impact of variation in $(D_{s_i})$ and $(\chi_i)$ on the expected demand of data consumers $(D_{c_i})$ and accordingly adjusts its price $(P_i)$.

\color{black} In Figure \ref{clod_payoff_externalities} \color{black}, we study the impact of varying the externality parameters $(\alpha \beta)$ and subsidizing factor $(\phi)$ on the cloud's payoff. As shown in the figure, the cloud platform obtains in general higher payoff when it follows the two-sided model, rather than the pay-as-to-go model. For example, under the two-sided model, the cloud platform receives \color{black}$1200$, $500$ and $400$ USD \color{black}as payoff when $(\alpha \beta)=0.2$ \color{black} and $\phi=5.0$, $2.0$ and $1.5$ \color{black}respectively. On the other hand, under the pay-as-to-go model and under the same externality and subsidizing parameters, the cloud platform receives less payoff of \color{black}$(200)$ USD. Similarly, data providers receive higher payoff and data consumers' demand is increased under the two-sided model compared to the pay-as-you-go model as shown respectively in \color{black}Figures \ref{data_provider_payoff_externalities} and \ref{data_consumer_demand_over_externalities} \color{black}.

In addition, we study the impact of the fifty-fity (egalitarian) choice on the two-sided market model, where the cloud platform and data providers share fifty percent of revenues considering the sitting of our two-sided model. As shown in Figure \ref{clod_payoff_externalities}, the fifty-fifty choice shows more efficient outcomes than the two-sided model in terms of cloud payoff under week externalities $(\alpha \beta \in [0.1-0.3])$ and certain values of subsidizing factor $\phi$. However, the cloud platform can receive higher payoff if it chooses higher subsidizing factor $\phi$ such as $\phi = 5$. Nevertheless, the egalitarian choice shows less efficient outcomes when the externalities become stronger (i.e $\alpha\beta \in [0.3 - 0.7]$) where the cloud platform receives less payoff under certain subsidizing factor such as $(\phi = 1.5)$ or $(\phi = 2)$. On the other hand, the fifty-fifty choice and the two-sided model show similar outcomes in terms of data providers' payoff and consumers demand as shown in Figures \ref{data_provider_payoff_externalities} and \ref{data_consumer_demand_over_externalities} respectively. However, as mentioned in Section \ref{sec:introduction}, the involved parties do not follow the egalitarian choice despite its good outcomes in some cases, mainly because of their greediness and because of high subsidies in real cloud markets. 

\color{black}

As can be observed from these figures, the two-sided market model shows less efficient outcomes as the externalities become stronger. Specifically, the cloud platform, data providers and consumers' surpluses gradually decrease as the values of the externality parameters increase. For example, in \color{black}Figure \ref{data_provider_payoff_externalities}, \color{black}the data providers receive a payoff of \color{black}$800$ USD \color{black} at an \color{black}externality parameters value of $(\alpha \beta = 0.2)$ and subsidizing factor of $\phi = 5.0$\color{black}, which is higher than that received with externality parameter value of $(\alpha \beta = 0.3)$ and subsidizing factor of \color{black}$\phi=5.0$\color{black}. This decrease continues until the two-sided market model reaches almost the same efficiency of the pay-to-go model under the strong externality values of \color{black}$[0.55$ - $0.7]$\color{black}. Similar behavior is observed in terms of cloud payoff and data consumers' demand as depicted in \color{black}Figures \ref{clod_payoff_externalities} and \ref{data_consumer_demand_over_externalities} \color{black}respectively. In fact, the cloud's payoff and data consumers' demand are higher under weak externality values, i.e., \color{black}$[0-0.5]$ \color{black} rather than in strong externality values \color{black}$[0.5-0.7]$ \color{black}. The reason behind such a behavior is that the cloud platform needs to provide more computing and storage resources under strong externalities to attract smaller number of data consumers, which adds more costs and then leads to less payoff\color{black}.

\begin{figure}[ht]
   \hspace{-0.5 cm}
    \includegraphics[scale=0.5]{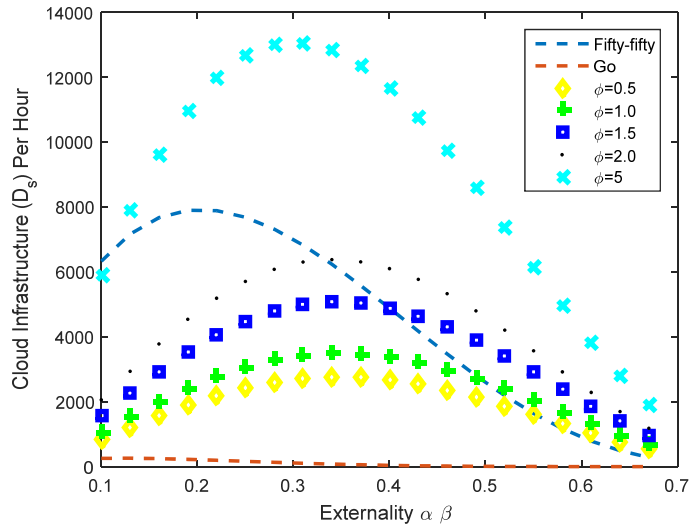}
   \caption{Cloud Infrastructure over externalities $\alpha \beta$}
    \label{Cloud_Infrastructure_over_externalities}
\end{figure}

Figure \ref{Cloud_Infrastructure_over_externalities} shows the number of computing infrastructure units provided by the cloud platform over varying externality values. As shown in the figure, at a subsidizing factor of $\phi = 2.0$ and externality values of $(\alpha \beta = 0.2)$ and $(\alpha \beta = 0.53)$\color{black}, the cloud platform provides $4800$ units of computing infrastructure. However, in Figure \ref{data_consumer_demand_over_externalities}, at a subsidizing factor of $(\phi = 2.0$) and externality values of $(\alpha \beta = 0.2)$, the cloud platform attracts $(3.3 \times 10^6)$ data consumers, while it attracts a smaller number of data consumers, i.e., $(0.5 \times 10^6)$ with externality values of $(\alpha \beta = 0.53)$\color{black}. In other words, the cloud platform attracts $(3.3 \times 10^6)$ of data consumers by providing $4800$ units of computing infrastructure at externality values of $(\alpha \beta = 0.2)$, while it attracts $(0.5 \times 10^6)$ data consumers by providing the same number of computing infrastructure units with externality values of $(\alpha \beta = 0.53)$\color{black}. Consequently, in such a case, the cloud platform asks for a higher portion $\chi_i$ of revenues to maximize its payoff, which negatively affects the payoff of the data providers. In Figure \ref{shared_revenue}, \color{black} we study the impact of the externality parameters on the portion of revenues asked by the cloud platform. As shown in the figure, the cloud platform asks for higher portions as the values of externality parameters are increased. For example, at a subsidizing factor of $\phi = 1.5$ and externality values of $(\alpha \beta = 0.3)$, data providers share $40\%$ of their revenues, while they share $68\%$ of their revenues with externalitiy values of $(\alpha \beta = 0.6)$ and a constant subsidizing factor of $\phi = 1.5$.


\subsection{Sensitivity analysis of subsidizing factor and greedy behavior of involved parties}\label{sensivity_anylsis_subsidizing_factor}

The exponential function $\chi_{i}^\phi$ captures the rational/greedy behavior of the cloud platform and data providers. The subsidizing factor $\phi$ implicitly describes the reactions of the cloud platform to the sharing portions offered by data providers. Theoretically, the cloud platform subsidizes the data providers by imposing a subsidizing factor $\phi$ that is less than $1$. This leads to having $\chi_{i}^\phi > \chi_i$ since the base $\chi_i$ is defined to be between $0$ and $1$, meaning that larger amounts of computing infrastructure units $D_{s_i}$ should be provided. On the other hand, data providers offer higher portions $\chi_i$ when the cloud platform acts greedily by imposing a subsidizing factor $\phi$ that is greater than $1$. This leads to having $\chi_{i}^\phi < \chi_{i}$, meaning that smaller amounts of computing infrastructure units $D_{s_i}$ need to be provided. When the subsidizing factor is equal to $1$, the data providers offer a market share to the cloud platform based on the size of its contribution. This leads to having $\chi_{i}^\phi = \chi_{i}$, meaning that the provided computing infrastructure is linearly probational with respect to the  market share $\chi_{i}$. In this case, neither the cloud platform nor data providers act greedily, and they do not subsidize each other at the same time.    \color{black}

In Figures \ref{clod_payoff_externalities}, \ref{data_provider_payoff_externalities} and \ref{data_consumer_demand_over_externalities}, we consider the impact of the subsidizing factor $\phi$ in the sensitivity analysis of the externalitiy parameters. As can be observed from those figures, the cloud platform and data consumers receive higher payoff than the data providers as the subsidizing factor increases. For example, \color{black} in Figure \ref{clod_payoff_externalities}\color{black}, the cloud's payoff is higher under a subsidizing factor of $\phi = 2.0$ than it is under a subsidizing factor of $\phi = 1.0$. Similarly, we can observe in Figure  \ref{data_consumer_demand_over_externalities} that data consumers' demand is higher under a subsidizing factor of $\phi = 2.0$ than it is under a a subsidizing factor of $\phi = 0.5$. Unlike the cases of cloud and data consumers, data providers' payoff is less under $\phi = 5.0$ than it is when $\phi = 1.5$, as depicted in Figure \ref{data_provider_payoff_externalities}.

\begin{figure}[!h]
    \centering
    \includegraphics[scale = 0.5]{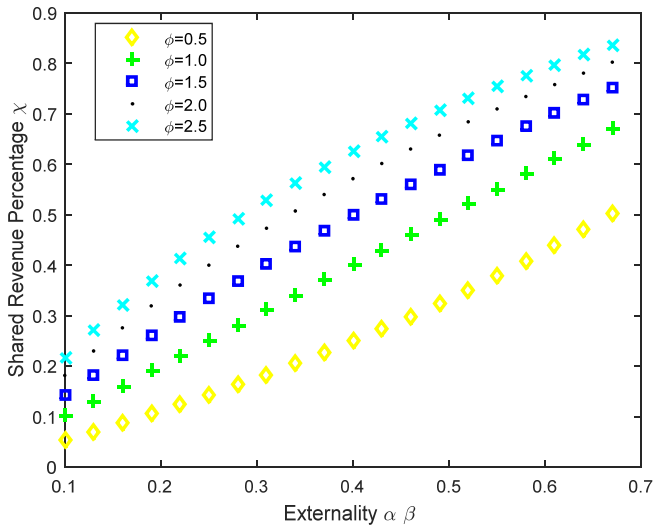}
    \caption{Shared revenue among the cloud and data providers $(\chi_i)$ over externalities $\alpha \beta$}
    \label{shared_revenue}
\end{figure}

\begin{figure}[!h]
    \centering
    \includegraphics[scale = 0.5]{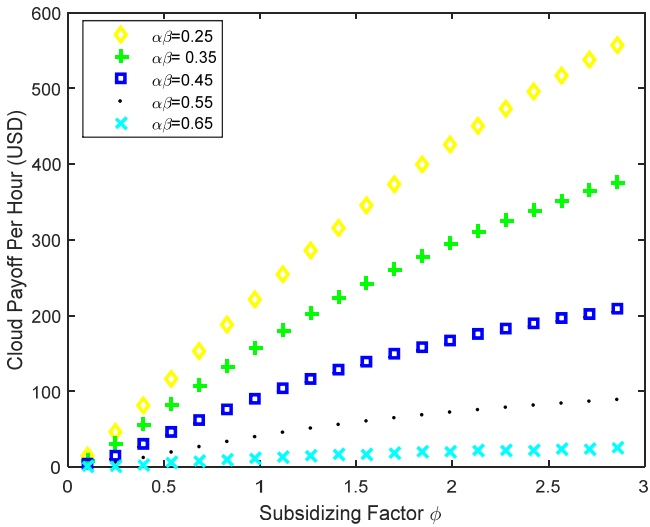}
    \caption{Cloud payoff over the subsidizing factor $\phi$}
    \label{cloud_payoff_over_phi}
\end{figure}
\begin{figure}[!h]
    \centering
    \includegraphics[scale = 0.5]{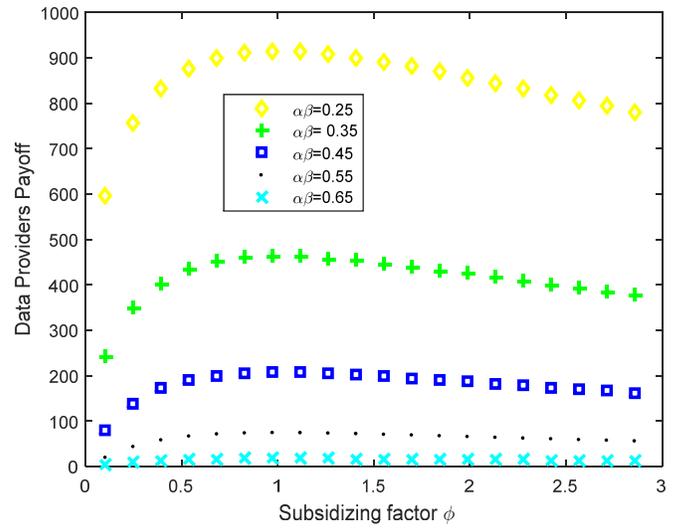}
    \caption{Data providers payoff over the subsidizing factor $\phi$}
    \label{data_provider_payoff_over_phi}
\end{figure}
\begin{figure}[!h]
    \centering
    \includegraphics[scale = 0.5]{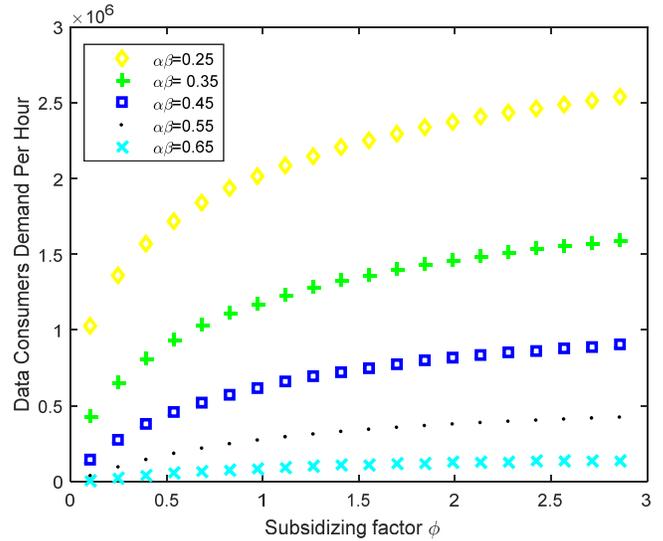}
    \caption{Consumer demand over the subsidizing factor $\phi$}
    \label{consumer_demand_over_phi}
\end{figure}

To better clarify the impact of the subsidizing factor on the payoffs of involved parties in our model, we run the simulation over a continuous, reasonable and wider range of subsidizing factor values. Specifically, we describe the payoff of the cloud, data providers and data consumers as a function of the subsidizing factor $\phi$ in Figures \ref{cloud_payoff_over_phi}, \ref{data_provider_payoff_over_phi} and \ref{consumer_demand_over_phi} respectively. The results shown in these figures confirm the insights extracted from Figures \ref{clod_payoff_externalities}, \ref{data_provider_payoff_externalities} and \ref{data_consumer_demand_over_externalities}. Furthermore, we notice in Figures \ref{data_provider_payoff_over_phi} and \ref{consumer_demand_over_phi} that the payoff of the data providers and consumers increases when the value of $\phi$ is between $0$ to $1$. Then slightly decreases$\ /$stabilizes as $\phi$ becomes greater than $1$. On the other hand, the cloud's payoff, as shown in Figure \ref{cloud_payoff_over_phi}, largely increases when $\phi$ becomes greater than $1$.

This behavior can be practically  interpreted with the results shown in Figure \ref{shared_revenue}. In this figure, we notice that data providers react to the greedy behavior of the cloud platform via increasing the shared revenue, which negatively affects the payoff of data providers as shown in Figures \ref{data_provider_payoff_externalities} and \ref{data_provider_payoff_over_phi}, but positively affects the cloud's payoff as shown in Figures \ref{clod_payoff_externalities} and \ref{cloud_payoff_over_phi}.

However, the behavior of the data providers (reacting to the greedy behavior of the cloud platform via increasing the shared portions) sustains a higher level of consumers' demand as shown in Figures \ref{cloud_payoff_over_phi} and \ref{data_consumer_demand_over_externalities}. Similarly, as shown in Figure \ref{shared_revenue}, the cloud platform acts to the low offered portion by imposing a subsidizing factor $\phi$ that is less than $1$. This negatively impacts the cloud's payoff but positively affects the cloud platforms' payoff. In summary, we conclude that our game includes a recovery mechanism that helps sustain efficient payoff outcomes for all the parties, in case any of the involved parties decides to act greedily toward the others.

\begin{figure}[H]
    \includegraphics[scale = 0.5]{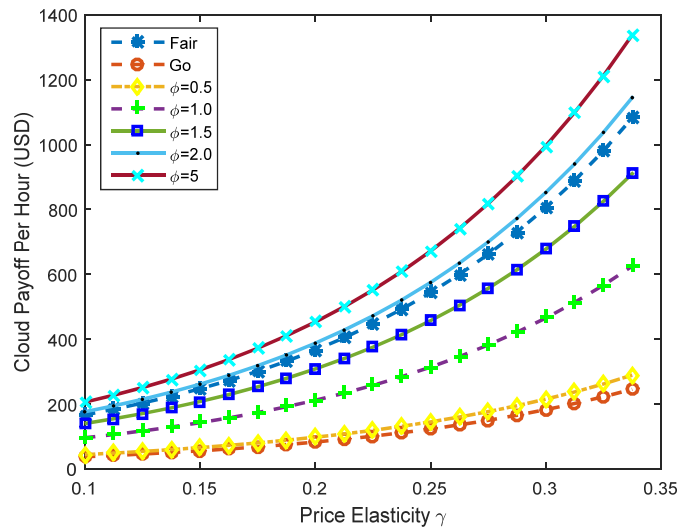}
    \caption{Cloud payoff over demand elasticity $\gamma$}
    \label{cloud_over_gamma}
\end{figure}

\subsection{Sensitivity analysis of consumer demands elasticity $(\gamma)$ and the multiplier ($k_1$)} \label{sensivity_anylsis_gamma}

\begin{figure}[!h]
    \includegraphics[scale =0.5]{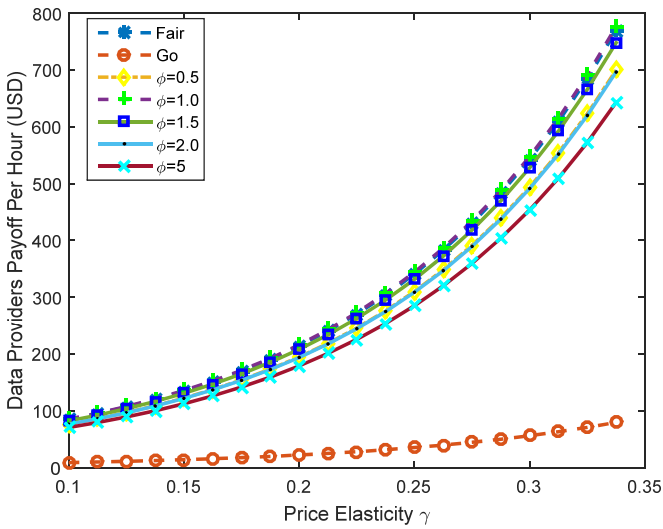}
    \caption{Data providers payoff over demand elasticity $\gamma$}
    \label{data_providers_over_gamma}
\end{figure}

\begin{figure}[!h]
    \includegraphics[scale =0.5]{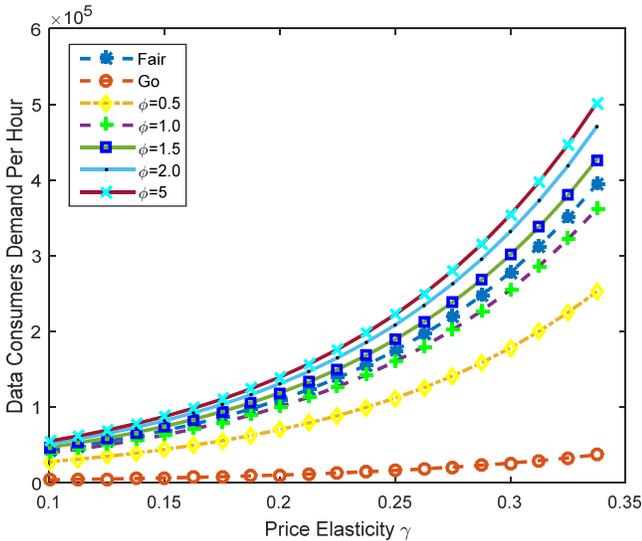}
    \caption{Consumers demand over demand elasticity $\gamma$}
    \label{data_consumers_over_gamma}
\end{figure}

We now move to analyzing the impact of consumers' demand elasticity on the surpluses of all involved parties (Figures \ref{cloud_over_gamma} - \ref{data_consumers_over_gamma}). A high negative elasticity value positively impacts the surplus of all parties under the two-sided game model through yielding higher payoffs for the cloud platform and data providers. The reason is that the exponential function $P_{i}^{-\gamma}$ increases when $\gamma$ increases from $0$ to $0.34$ as long as the base $P_i$ is less than $1$. This makes data providers able to charge their consumers higher prices without significantly leading to a decrease in the demands.

\begin{figure}[!h]
    \includegraphics[scale = 0.5]{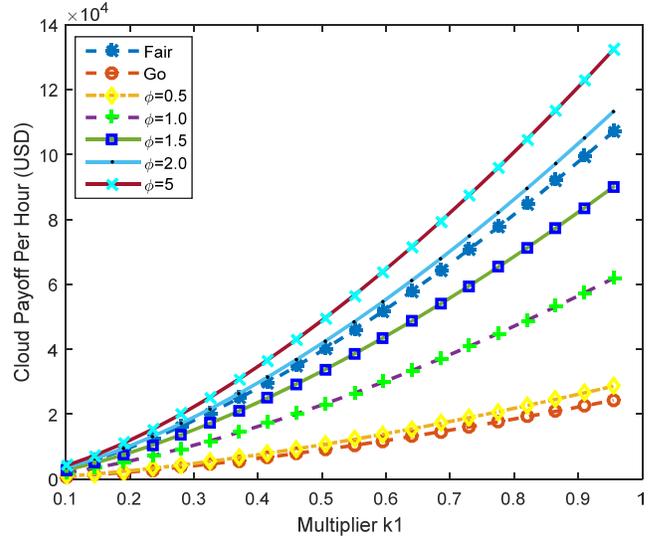}
    \caption{Cloud payoff over multiplier $k_1$}
    \label{cloud_over_k1}
\end{figure}

\begin{figure}[!h]
    \includegraphics[scale = 0.5]{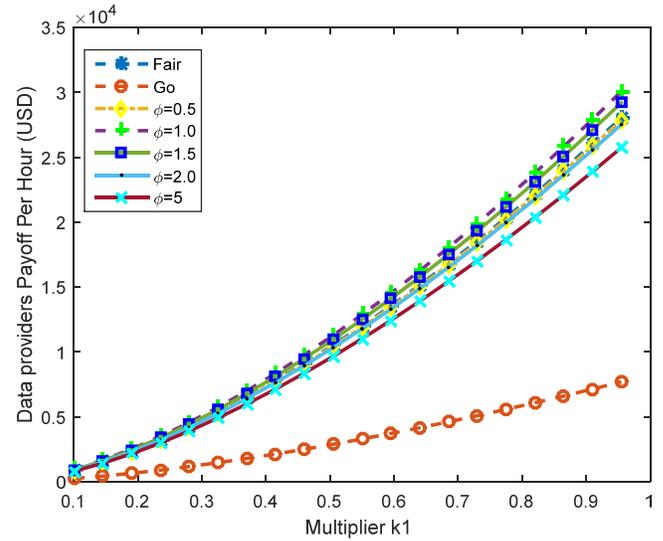}
    \caption{Data providers payoff over multiplier $k_1$}
    \label{data_provider_over_k1}
\end{figure}


\begin{figure}[H]
    \includegraphics[scale = 0.5]{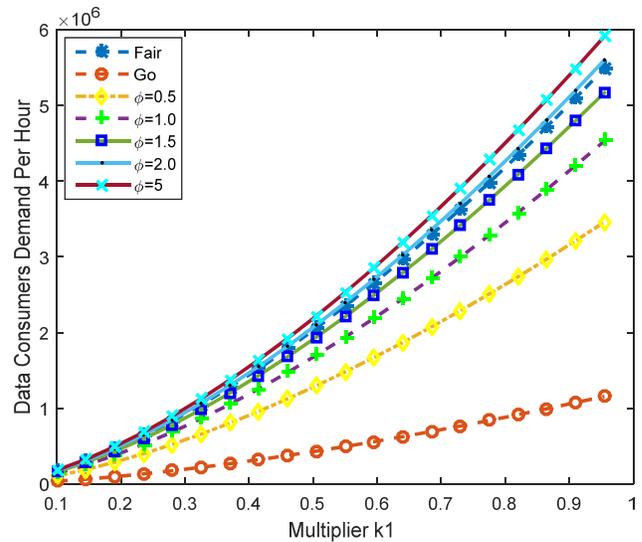}
    \caption{Consumers demand over multiplier $k_1$}
    \label{dc_over_k1}
\end{figure}

We also study the impact of the multiplier $k_1$ on the surplus of all involved parties. As shown in Figures \ref{cloud_over_k1} - \ref{dc_over_k1}, the surpluses of the cloud platform, service data providers and consumers increase as the value of the multiplier $k_1$ increases. 
The reason is that increasing the multiplier $k_1$ leads to increasing consumers' demands, which positively affects the total revenues.

\section{Conclusion}\label{sec:Conclusion}

In this paper, we proposed a game theoretical model based on the two-sided market theory to monetize big data services over the cloud. In particular, we studied the impact of cross group externalities between the amount of provided cloud resources and number of service consumers. The objective is to come up with a new vision in which the cloud can play a primordial role in introducing big data service providers and data consumers to each other, which results in higher benefits for all the involved players. To achieve this goal, we designed a game theoretical model in which the cloud platform and big data service providers engage in a closed loop of dependencies that makes them interested in satisfying each other interest, instead of following an aggressive competition strategy. 
Empirical results showed that our model outperforms the state-of-the-art cloud business models, i.e., the fifty-fifty and pay-as-you-go models in terms of total surpluses earned by the different parties.


\bibliographystyle{IEEEtran}
\footnotesize
\bibliography{IEEEabrv,ref}

\begin{thebibliography}{10}
\providecommand{\url}[1]{#1}
\csname url@samestyle\endcsname
\providecommand{\newblock}{\relax}
\providecommand{\bibinfo}[2]{#2}
\providecommand{\BIBentrySTDinterwordspacing}{\spaceskip=0pt\relax}
\providecommand{\BIBentryALTinterwordstretchfactor}{4}
\providecommand{\BIBentryALTinterwordspacing}{\spaceskip=\fontdimen2\font plus
\BIBentryALTinterwordstretchfactor\fontdimen3\font minus
  \fontdimen4\font\relax}
\providecommand{\BIBforeignlanguage}[2]{{%
\expandafter\ifx\csname l@#1\endcsname\relax
\typeout{** WARNING: IEEEtran.bst: No hyphenation pattern has been}%
\typeout{** loaded for the language `#1'. Using the pattern for}%
\typeout{** the default language instead.}%
\else
\language=\csname l@#1\endcsname
\fi
#2}}
\providecommand{\BIBdecl}{\relax}
\BIBdecl

\bibitem{DMRAMAZON2018}
``Dmr amazon statistical report 2018,''
  \url{https://expandedramblings.com/index.php/downloads/dmr-amazon-web-services-report/},
  accessed: 2019-01-31.

\bibitem{Niyato2018}
Y.~Jiao, P.~Wang, S.~Feng, and D.~Niyato, ``Profit maximization mechanism and
  data management for data analytics services,'' \emph{IEEE Internet of Things
  Journal}, vol.~5, no.~3, pp. 2001--2014, 2018.

\bibitem{Niyato2016}
D.~Niyato, D.~T. Hoang, N.~C. Luong, P.~Wang, D.~I. Kim, and Z.~Han, ``Smart
  data pricing models for the internet of things: a bundling strategy
  approach,'' \emph{IEEE Network}, vol.~30, no.~2, pp. 18--25, 2016.

\bibitem{Niyato20166}
A.~Jin, W.~Song, P.~Wang, D.~Niyato, and P.~Ju, ``Auction mechanisms toward
  efficient resource sharing for cloudlets in mobile cloud computing,''
  \emph{IEEE Transactions on Services Computing}, vol.~9, no.~6, pp. 895--909,
  2016.

\bibitem{BATAINEH2019}
A.~S. Bataineh, R.~Mizouni, J.~Bentahar, and M.~{El Barachi}, ``Toward
  monetizing personal data: A two-sided market analysis,'' \emph{Future
  Generation Computer Systems}, vol. 111, pp. 435--459, 2020.

\bibitem{Chakareski2015}
J.~Chakareski, ``Cost and profit driven cloud-{P2P} interaction,''
  \emph{Peer-to-Peer Networking and Applications}, vol.~8, no.~2, pp. 244--259,
  Mar 2015.

\bibitem{Zehui2017}
Z.~Xiong, S.~Feng, D.~Niyato, P.~Wang, and Y.~Zhang, ``Competition and
  cooperation analysis for data sponsored market: {A} network effects model,''
  \emph{CoRR}, vol. abs/1711.01054, 2017.

\bibitem{GUTH1982367}
W.~Güth, R.~Schmittberger, and B.~Schwarze, ``An experimental analysis of
  ultimatum bargaining,'' \emph{Journal of Economic Behavior and Organization},
  vol.~3, no.~4, pp. 367 -- 388, 1982.

\bibitem{Rochet2003}
J.~Rochet and J.~Tirole, ``Platform competition in two-sided markets,''
  \emph{Journal of the European Economic Association}, vol.~1, no.~4, pp.
  990--1029, 2003.

\bibitem{Greenberg2008}
A.~Greenberg, J.~Hamilton, D.~A. Maltz, and P.~Patel, ``The cost of a cloud:
  Research problems in data center networks,'' \emph{SIGCOMM Comput. Commun.
  Rev.}, vol.~39, no.~1, pp. 68--73, Dec. 2008.

\bibitem{Zhang2015}
N.~Zhang and H.~Hämmäinen, ``Cost efficiency of {SDN} in {LTE}-based mobile
  networks: Case {Finland},'' in \emph{2015 International Conference and
  Workshops on Networked Systems (NetSys)}, 2015, pp. 1--5.

\bibitem{Rebai2015}
S.~Rebai, M.~Hadji, and D.~Zeghlache, ``Improving profit through cloud
  federation,'' in \emph{2015 12th Annual IEEE Consumer Communications and
  Networking Conference (CCNC)}, 2015, pp. 732--739.

\bibitem{Lin2015}
Y.~Lin and H.~Shen, ``Autotune: game-based adaptive bitrate streaming in
  {P2P}-assisted cloud-based {VoD} systems,'' in \emph{2015 IEEE International
  Conference on Peer-to-Peer Computing (P2P)}, 2015, pp. 1--10.

\bibitem{Ranjan2013}
R.~Pal and P.~Hui, ``Economic models for cloud service markets: Pricing and
  capacity planning,'' \emph{Theoretical Computer Science}, vol. 496, pp. 113
  -- 124, 2013.

\bibitem{Xu2012}
H.~Xu and B.~Li, ``A general and practical datacenter selection framework for
  cloud services,'' in \emph{2012 IEEE Fifth International Conference on Cloud
  Computing}, 2012, pp. 9--16.

\bibitem{Ding2015}
J.~Ding, R.~Yu, Y.~Zhang, S.~Gjessing, and D.~H.~K. Tsang, ``Service provider
  competition and cooperation in cloud-based software defined wireless
  networks,'' \emph{IEEE Communications Magazine}, vol.~53, no.~11, pp.
  134--140, 2015.

\bibitem{Samimi2011}
P.~Samimi and A.~Patel, ``Review of pricing models for \& cloud computing,'' in
  \emph{2011 IEEE Symposium on Computers \& Informatics}, 2011, pp. 634--639.

\bibitem{wahab2021federated}
O.~A. Wahab, A.~Mourad, H.~Otrok, and T.~Taleb, ``Federated machine learning:
  Survey, multi-level classification, desirable criteria and future directions
  in communication and networking systems,'' \emph{IEEE Communications Surveys
  \& Tutorials}, 2021.

\bibitem{Valerio2013}
V.~D. Valerio, V.~Cardellini, and F.~L. Presti, ``Optimal pricing and service
  provisioning strategies in cloud systems: {A} stackelberg game approach,'' in
  \emph{{IEEE} {CLOUD}}.\hskip 1em plus 0.5em minus 0.4em\relax {IEEE} Computer
  Society, 2013, pp. 115--122.

\bibitem{Luong2017}
N.~C. Luong, P.~Wang, D.~Niyato, Y.~Wen, and Z.~Han, ``Resource management in
  cloud networking using economic analysis and pricing models: A survey,''
  \emph{IEEE Communications Surveys Tutorials}, vol.~19, no.~2, pp. 954--1001,
  2017.

\bibitem{Luong2016}
N.~C. Luong, D.~T. Hoang, P.~Wang, D.~Niyato, D.~I. Kim, and Z.~Han, ``Data
  collection and wireless communication in internet of things {(IoT)} using
  economic analysis and pricing models: A survey,'' \emph{IEEE Communications
  Surveys Tutorials}, vol.~18, no.~4, pp. 2546--2590, 2016.

\bibitem{Niyato2020}
Y.~{Zhang}, Z.~{Xiong}, D.~{Niyato}, P.~{Wang}, H.~V. {Poor}, and D.~I. {Kim},
  ``A game-theoretic analysis for complementary and substitutable {IoT}
  services delivery with externalities,'' \emph{IEEE Transactions on
  Communications}, vol.~68, no.~1, pp. 615--629, 2020.

\bibitem{Bataineh2016}
A.~S. Bataineh, R.~Mizouni, M.~E. Barachi, and J.~Bentahar, ``Monetizing
  personal data: A two-sided market approach,'' \emph{Procedia Computer
  Science}, vol.~83, pp. 472 -- 479, 2016.

\bibitem{Mashayekhy2014ATM}
L.~Mashayekhy, M.~M. Nejad, and D.~Grosu, ``A two-sided market mechanism for
  trading big data computing commodities,'' \emph{2014 IEEE International
  Conference on Big Data (Big Data)}, pp. 153--158, 2014.

\bibitem{wahab2016stackelberg}
O.~A. Wahab, J.~Bentahar, H.~Otrok, and A.~Mourad, ``A stackelberg game for
  distributed formation of business-driven services communities,'' \emph{Expert
  Systems with Applications}, vol.~45, pp. 359--372, 2016.

\bibitem{google2019}
Google, ``Google clustred data,'' \url{https://github.com/google/cluster-data},
  [Online; accessed 19-July-2019].

\bibitem{Amazon_Calculator}
Amazon, ``Simple monthly calculator,''
  \url{https://calculator.s3.amazonaws.com/index.html}, [Online; accessed
  19-July-2019].

\bibitem{Amazon_marketplace}
------, ``{IoT} and big data services in {Amazon} market places,''
  \url{https://aws.amazon.com/marketplace/search?page=1&category=96c2cd16-fe69-4b18-99cc-e016c61e820c},
  [Online; accessed 19-Nov-2019].

\bibitem{Danaher2002}
P.~J. Danaher, ``Optimal pricing of new subscription services: Analysis of a
  market experiment,'' \emph{Marketing Science}, vol.~21, no.~2, pp. 119--138,
  2002.

\end{thebibliography}

\vspace{10mm}

\begin{wrapfigure}[9]{l}{25mm}
\vspace{-5mm}
    \includegraphics[width=1in,height=1.45in,clip, keepaspectratio]{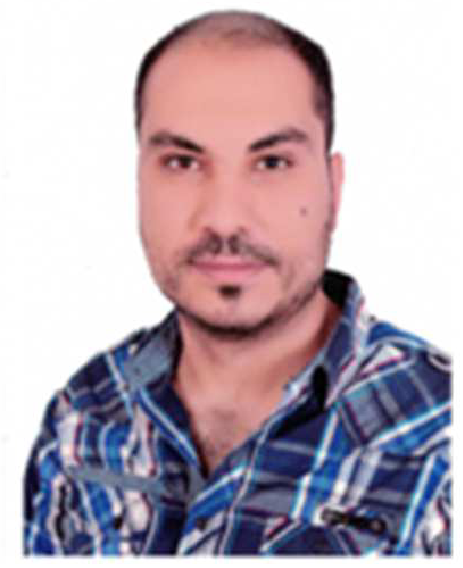} 

  \end{wrapfigure}\par

   \textbf {Ahmed Saleh Bataineh} received the bachelor’s degree in computer engineering from Jordan University of science and technology, Jordan, in 2010, and the master's degree in electrical and computer engineering from Concordia University, Canada, in 2014. He is currently pursuing the Ph.D. degree with the Concordia Institute for Information Systems Engineering (CIISE) under the supervision of Prof. Jamal Bentahar. His past research activities include multi-agent systems, verification and model checking. His current research interests include economics, optimization problems and game theory applied to IoT services and big data.
	\par

\vspace{5mm}
	\begin{wrapfigure}[9]{l}{25mm}
\vspace{-5.5mm}
    \includegraphics[width=1in,height=1.45in,clip, keepaspectratio]{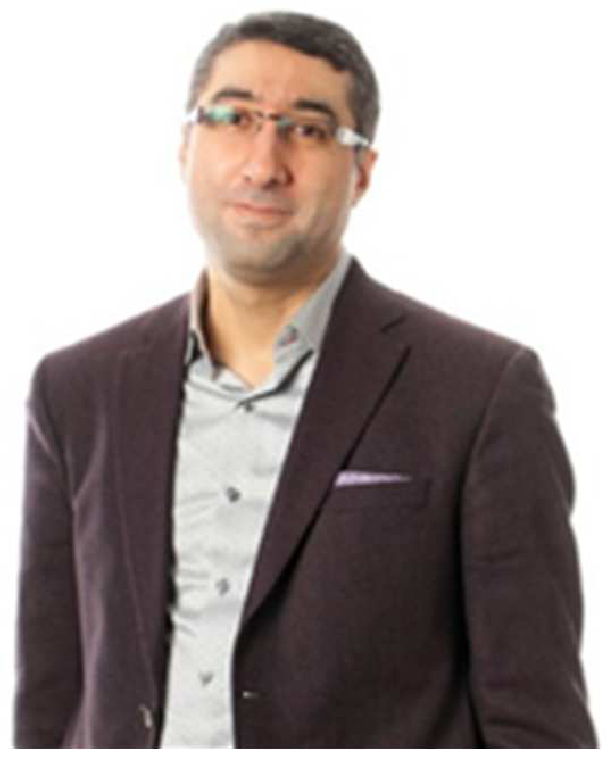} 
  \end{wrapfigure}\par

   \textbf {Jamal Bentahar} received the Ph.D. degree in computer science and software engineering from Laval University, Canada, in 2005. He is a Professor with Concordia Institute for Information Systems Engineering, Concordia University, Canada. From 2005 to 2006, he was a Postdoctoral Fellow with Laval University, and then NSERC Postdoctoral Fellow at Simon Fraser University, Canada. He is an NSERC Co-Chair for Discovery Grant for Computer Science (2016-2018). His research interests include the areas of computational logics, model checking, multi-agent systems, service computing, game theory, and software engineering.
	\par

\vspace{5mm}

\begin{wrapfigure}[7]{l}{25mm}
\vspace{-5mm}
    \includegraphics[width=1.0in,height=0.9in]{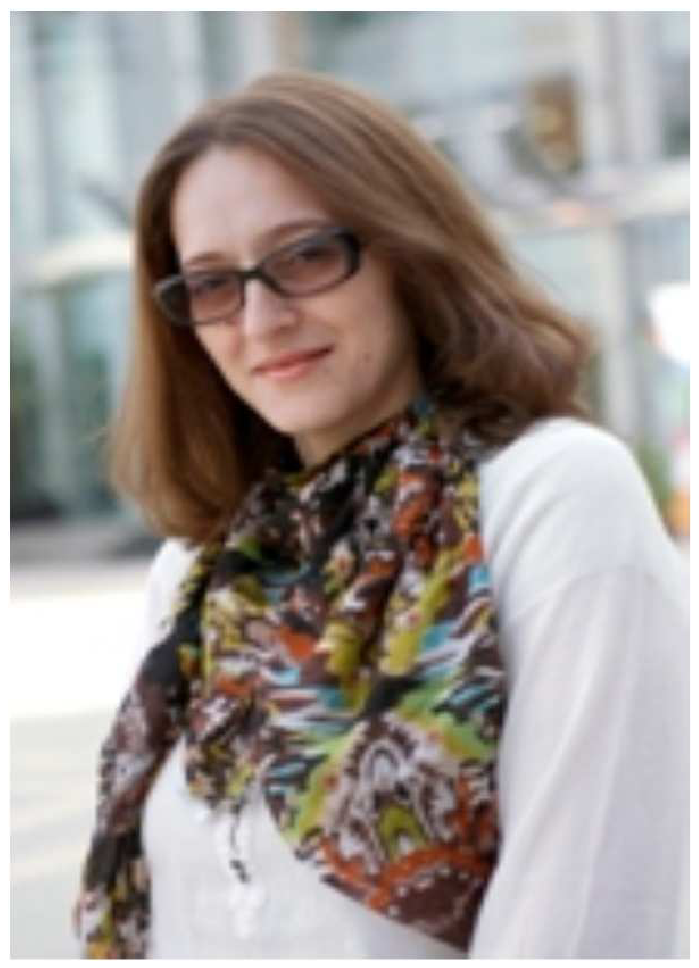} 
  \end{wrapfigure}\par

   \textbf{Rabeb Mizouni} is an associate professor in Electrical and Computer Engineering at Khalifa University. She got her PhD and her MSc in Electrical and Computer Engineering from Concordia University, Montreal, Canada in 2007 and 2002 respectively. Currently, she is interested in the deployment of context aware mobile applications, software product line and cloud computing.
	\par
	
\vspace{8mm}
	
\begin{wrapfigure}[8]{l}{27mm}
\vspace{-8mm}
    \includegraphics[width=1in,height=1.45in,clip, keepaspectratio]{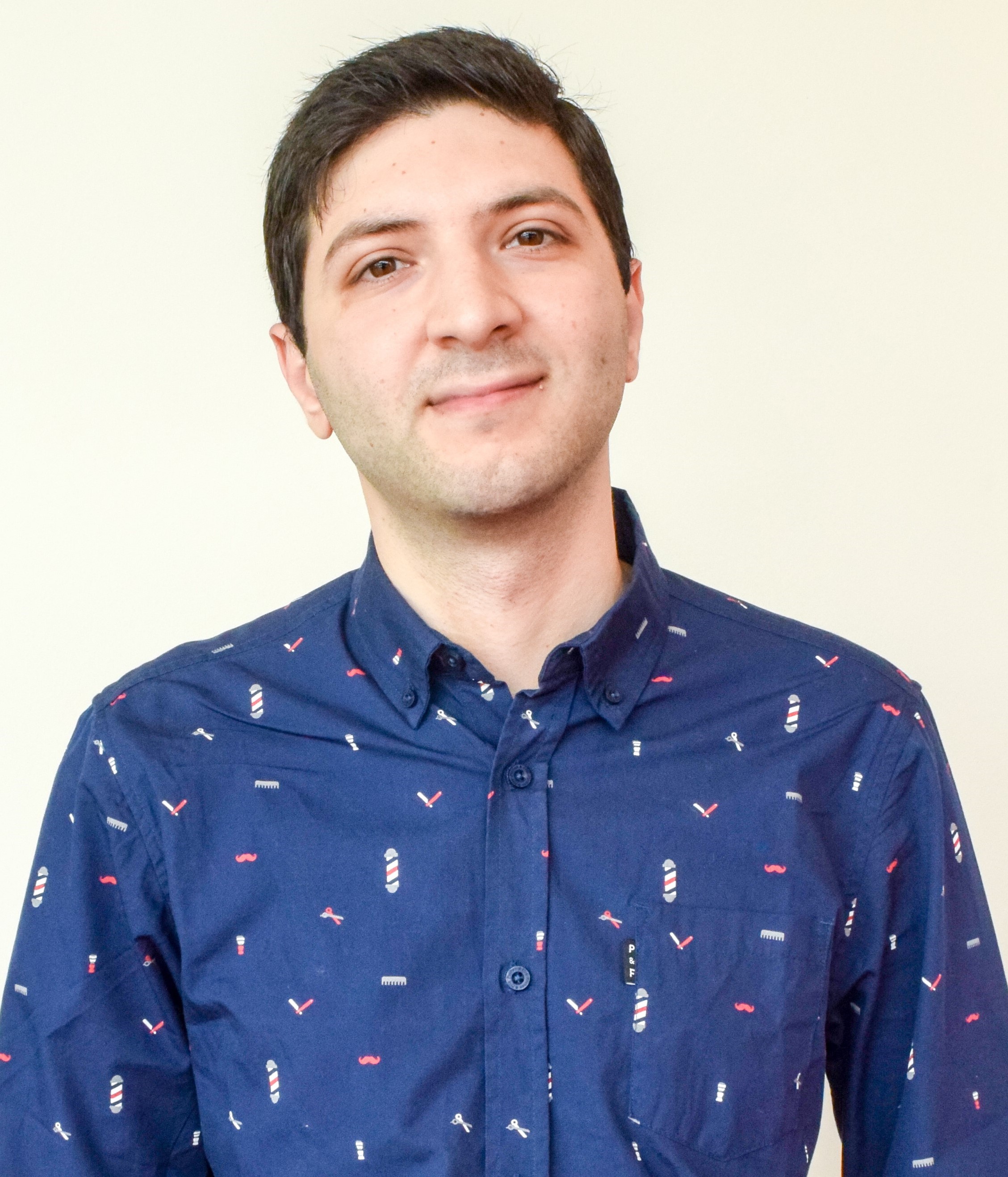} 
  \end{wrapfigure}\par

  \textbf{Omar Abdel Wahab} is an Assistant Professor at the Department of Computer Science and Engineering, Université du Québec en Outaouais, Canada. He holds a Ph.D. in Information and Systems Engineering from Concordia University, Montreal, Canada. He received
his M.Sc. in computer science in 2013 from the Lebanese American University (LAU), Lebanon. From 2017 to 2018, he was a postdoctoral fellow at the École de Technologie Supérieure (ÉTS), Canada, where he worked on an industrial research project in collaboration with Rogers and Ericsson. The main topics of his current research activities are in the areas of cybersecurity, data analytics, Internet of Things (IoT) and game theory.
\par

\vspace{8mm}

\begin{wrapfigure}[7]{l}{27mm}
\vspace{-4mm}
    \includegraphics[width=1.1in,height=1.45 in,clip, keepaspectratio]{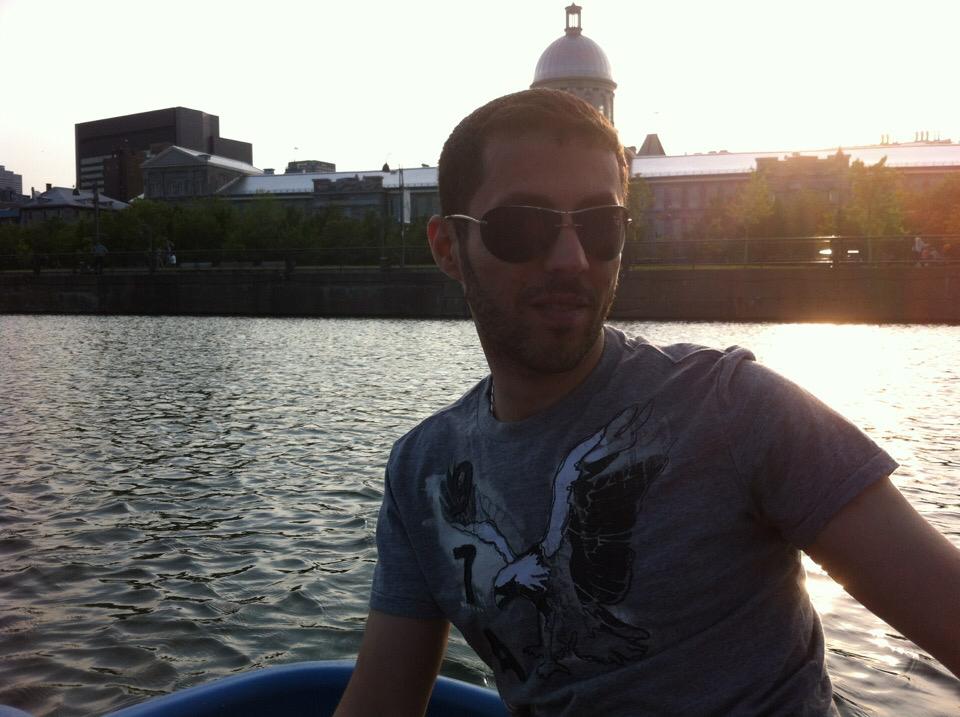}
  \end{wrapfigure}\par

\textbf{Gaith Rjoub} received the B.S. degree in information technology engineering from Arab Open University, Amman, Jordan, in 2008  and M.Eng. degree in Quality Systems Engineering from Concordia Institute for Information Systems Engineering (CIISE), Canada in 2014. He is currently pursuing his Ph.D. degree in Information Systems Engineering at concordia University. His research interests include cloud computing, machine learning, artificial intelligence and big data analytics. \par

\vspace{5mm}
	
\begin{wrapfigure}[9]{l}{27mm}
\vspace{-5.5mm}
    \includegraphics[width=1in,height=1.45in,clip, keepaspectratio]{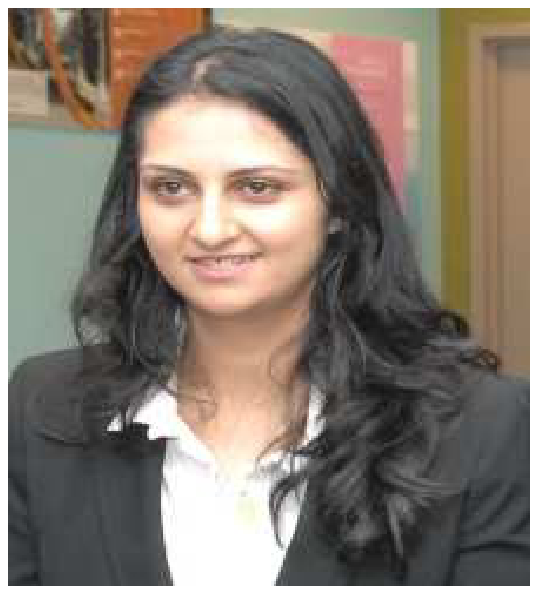} 
  \end{wrapfigure}\par
	
	\textbf{May El Barachi} is a next generation networking expert holding Ph.D. and master’s degrees in Computer Engineering from Concordia University (Canada). She has 12 years’ experience in the field with a strong focus on smart and resource efficient systems. During those years, she acquired academic experience as associate professor and has worked as researcher with Ericsson Research Canada. She co-founded and directed a cutting-edge research lab aiming at investigating sensory, cloud, and mobile technologies. Presently, she works as Associate Professor at University of Wollongong Dubai and is involved in several research collaborations with international universities and local industrial partners. \par

\end{document}